\documentclass[sigconf]{acmart}

\acmYear{2019}
\acmISBN{} 
\acmDOI{} 
\startPage{1}

\setcopyright{none}

\usepackage{booktabs}   
\usepackage{subcaption} 

\usepackage{footnote}
\usepackage{amsmath,amssymb,amsfonts}
\usepackage{graphicx}
\usepackage{textcomp}
\usepackage{algorithm}
\usepackage{algorithmic}
\usepackage{xcolor}
\usepackage{threeparttable}

\usepackage{amsthm}
\usepackage{mathtools}
\usepackage{xparse}
\NewDocumentCommand{\codeword}{v}{%
\texttt{\textcolor{blue}{#1}}%
}

\usepackage{comment}
\def\BibTeX{{\rm B\kern-.05em{\sc i\kern-.025em b}\kern-.08em
    T\kern-.1667em\lower.7ex\hbox{E}\kern-.125emX}}
\usepackage{listings}
\definecolor{mygreen}{rgb}{0,0.6,0}
\definecolor{mygray}{rgb}{0.5,0.5,0.5}
\definecolor{mymauve}{rgb}{0.58,0,0.82}
\definecolor{navy}{RGB}{0,0,128}
\definecolor{cornflowerblue}{RGB}{100, 149, 237}
\definecolor{lightgoldenrod}{RGB}{238, 221, 130}
\definecolor{goldenrod}{RGB}{218, 165, 32}
\definecolor{tomato}{RGB}{255, 99, 71}
\definecolor{orangered}{RGB}{255, 69, 0}
\definecolor{lightcoral}{RGB}{240, 128, 128}
\definecolor{lightseagreen}{RGB}{32, 178, 170}
\definecolor{yellowgreen}{RGB}{154, 205, 50}
\definecolor{skyblue}{RGB}{135, 206, 250}
\definecolor{mistyrose}{RGB}{255, 228, 225}
\definecolor{bananamania}{rgb}{0.98, 0.91, 0.71}
\definecolor{blond}{rgb}{0.98, 0.94, 0.75}
\definecolor{champagne}{rgb}{0.97, 0.97, 0.81}
\definecolor{coralpink}{rgb}{0.97, 0.51, 0.47}

\usepackage{xcolor}
\usepackage{multirow}
\usepackage{tabularx}
\usepackage{graphics}
\usepackage{amsmath}
\usepackage{proof, environ, array}
\usepackage{tikz}
\newif\ifcenterinfermeasure
\NewEnviron{infertable}
 {\centerinfermeasuretrue
  \setbox0=\vbox{tabskip=0pt
    \renewcommand\\[1][]{\cr}%
    \halign{##&##\cr
    \BODY\crcr}
    \setbox0=\lastbox
    \setbox0=\hbox{\unhbox0 \unskip\setbox2=\lastbox\unskip\setbox4=\lastbox
      \global\dimen1=\wd4 \global\dimen3=\wd2 }
  }%
  \centerinfermeasurefalse
  \begin{tabular}{%
    @{}
    >{\centering$\displaystyle}p{\dimen1}<{$}
    >{\centering\arraybackslash$\displaystyle}p{\dimen3}<{$}
    @{}
  }<
  \BODY
  \end{tabular}
}
\usepackage{stackengine}
\def\eqgap{.2ex}
\def\overgap{.4ex}
\def\inferrulerule{.5pt}

\newlength\rulelength
\newlength\toplength
\newlength\bottomlength

\newcommand\myinferrule[2]{%
  \stackMath%
  \setlength\bottomlength{\widthof{$#1$}}%
  \setlength\toplength{\widthof{$#2$}}%
  \ifdim\toplength>\bottomlength%
    \setlength\rulelength{\the\toplength}%
  \else%
    \setlength\rulelength{\the\bottomlength}%
  \fi%
  \mathrel{%
    \stackunder[\overgap]{%
      \stackon[\overgap]{%
        \stackanchor[\eqgap]%
          {\rule{\the\rulelength}{\inferrulerule}}%
        {\rule{\the\rulelength}{\inferrulerule}}%
      }{#2}%
    }{#1}%
  }%
}

\newcommand{\inlinecode}[2]{\colorbox{white}{\lstinline[language=#1]$#2$}}
\newcommand{\ignore}[1]{}

\usepackage{color}
\definecolor{atomictangerine}{rgb}{1.0, 0.6, 0.4}
\definecolor{antiquefuchsia}{rgb}{0.57, 0.36, 0.51}
\definecolor{pastelorange}{rgb}{1.0, 0.7, 0.28}
 
\usepackage{amssymb}
\usepackage{pifont}
\newcommand{\cmark}{\ding{51}}%
\newcommand{\xmark}{\ding{55}}%

\usepackage{pict2e,xcolor}

\usepackage{ragged2e}

\usepackage{url}
\usepackage{hyperref} 

\newcommand{\hdsenDouble}[1]{\texttt{SEN\_{HD$_D$}}}  
\newcommand{\hdsenSingle}[1]{\texttt{SEN\_{HD$_S$}}} 

\newcommand{\UKD}{\mathtt{UKD}}  
\newcommand{\SID}{\mathtt{SID}}  
\newcommand{\RUD}{\mathtt{RUD}}  

\newcommand{\SECRET}{\mathit{SECRET}}

\newtheorem{thm}{Theorem}[section]
\newtheorem{defn}[thm]{Definition}

\begin{document}
\title{Mitigating Power Side Channels during Compilation}

\author{Jingbo Wang}
\affiliation{
  \institution{University of Southern California}   
  \city{Los Angeles}
  \state{California}
  \postcode{90089}
  \country{USA}                
}

\author{Chungha Sung}
\affiliation{
  \institution{University of Southern California}   
  \city{Los Angeles}
  \state{California}
  \postcode{90089}
  \country{USA}                
}

\author{Chao Wang}
\affiliation{
  \institution{University of Southern California}   
  \city{Los Angeles}
  \state{California}
  \postcode{90089}
  \country{USA}                
}

\begin{abstract}
The code generation modules inside modern compilers such as GCC and
LLVM, which use a limited number of CPU registers to store a large
number of program variables, may introduce side-channel leaks even in
software equipped with state-of-the-art countermeasures.  We propose a
program analysis and transformation based method to eliminate this
side channel.  Our method has a type-based technique for detecting
leaks, which leverages Datalog-based declarative analysis and
domain-specific optimizations to achieve high efficiency and accuracy.
It also has a mitigation technique for the compiler's backend, more
specifically the register allocation modules, to ensure that
potentially leaky intermediate computation results are always stored
in different CPU registers or spilled to memory with isolation.
We have implemented and evaluated our method in LLVM for the x86
instruction set architecture.  Our experiments on cryptographic
software show that the method is effective in removing the side
channel while being efficient, i.e., our mitigated code is more
compact and runs faster than code mitigated using state-of-the-art
techniques.
\end{abstract}

\ignore{



}

\maketitle

\section{Introduction}
\label{sec:introduction}

Cryptography is an integral part of many security protocols, which in
turn are used by numerous applications.  However, despite the strong
theoretical guarantee, cryptosystems in practice are vulnerable to
side-channel attacks when non-functional properties such as timing,
power and electromagnetic radiation are exploited to gain
information about sensitive data~\cite{kocher1996timing,
chari1999towards, messerges1999investigations,
clavier2000differential, quisquater2001electromagnetic,
messerges2002examining, brier2004correlation, zhou2005side,
standaert2009unified, hund2013practical}.  For example, if the power
consumption of a device running an encryption algorithm depends on the
secret key, statistical techniques such as differential power analysis
(DPA) can be used to perform attacks
reliably~\cite{kocher1999differential, clavier2000differential,
brier2004correlation, messerges2000using,moradi2014side}.

Although there are techniques for mitigating power side
channels~\cite{zhang2018scinfer,eldib2014synthesis,eldib2014formal,
bhunia2014hardware,almeida2013formal,bayrak2013sleuth,akkar2003generic},
they focus exclusively on the \emph{Boolean level}, e.g., by
targeting circuits in cryptographic hardware or software code
that has been converted to bit-level representations~\cite{IshaiSW03}.
This limits the use of such techniques in real compilers; as a result,
none of them was able to fit into modern compilers such as GCC and
LLVM to directly handle the
\emph{word-level} intermediate representation (IR). 
In addition, code transformations in compilers may add new side
channels, even if the input program is equipped with state-of-the-art
countermeasures.

Specifically, compilers tend to use a limited number of the CPU's
registers to store a potentially-large number of intermediate
computation results of a program.  And, when two masked and hence
de-sensitized values are put into the same register, it is possible
for the \emph{masking} countermeasure to be removed accidentally.
We will show, as part of this work, that even provably-secure
techniques such as high-order
masking~\cite{barthe2015verified,barthe2016strong,balasch2014cost} is vulnerable to
such leaks.  Indeed, we have found leaks in the compiled code
produced by LLVM for both x86 and MIPS/ARM platforms, regardless of
whether the input program is equipped with high-order masking.

To solve the problem, we propose a secure compilation method with two
main contributions.
First, we introduce a type-inference system to soundly and quickly
detect power side-channel leaks.  By soundly, we mean that the system
is conservative and guarantees not to miss real leaks. By quickly, we
mean that it relies only on syntactic information of the program and
thus can be orders-of-magnitude faster than formal
verification~\cite{eldib2014formal,zhang2018scinfer}.
Second, we propose a mitigation technique for the compiler's backend
modules to ensure that, for each pair of intermediate variables that
may cause side-channel leaks, the two values are always stored in
different registers or memory locations.

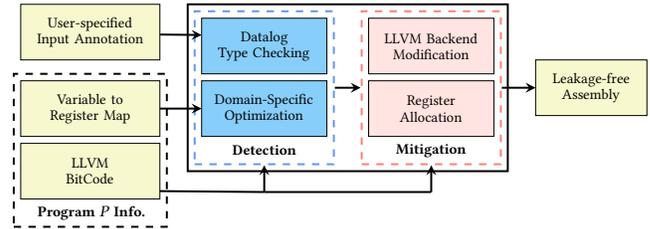
\begin{figure}
\vspace{2ex}
\centering
\scalebox{0.925}{\begin{tikzpicture}[font=\scriptsize] 
  \tikzstyle{arrow}=[thick,->,>=stealth,black]
  \tikzstyle{txt}=[above=5pt,right,text width=1.6cm]
  \tikzstyle{long_txt}=[above=5pt,right,text width=2cm]
  \tikzstyle{short_txt}=[above=5pt,right,text width=1.4cm]
    	
  	
    \node[draw=black, fill=champagne, minimum width=20mm, minimum height=8mm, text width=20mm, align=center, inner sep=0,outer sep=0] (U0) at (0.0, 3.1) {User-specified\\ Input Annotation};

    \node[draw=black, fill=champagne, minimum width=20mm, minimum height=8mm, text width=20mm, align=center, inner sep=0,outer sep=0] (S0) at (0.0, 2.0) {Variable to  \\  Register Map};
    
    \node[draw=black, fill=champagne, minimum width=20mm, minimum height=8mm, text width=20mm, align=center, inner sep=0,outer sep=0] (P1) at (0.0, 1.1) {LLVM \\ BitCode};
    
    \node[minimum width=22mm, minimum height=22mm, dashed, draw=black, thick, align=center] (PR) at (0.0, 1.4) {};    
    \node[text width=1.6cm, align=center] at (0.0, 0.5) {\textbf{Program $P$ Info.}};

    \node[] (D0) at (2.5, 0.8) {};
    
    \node[] (D1) at (4.9, 0.8) {};

    \node[draw=black, fill=skyblue, minimum width=18mm, minimum height=8mm, text width=18mm, align=center, inner sep=0,outer sep=0] (K0) at (2.5, 2.9) {Datalog \\ Type Checking};
    \node[draw=black, fill=skyblue, minimum width=18mm, minimum height=8mm, text width=18mm, align=center, inner sep=0,outer sep=0] (K1) at (2.5, 2.0) {Domain-Specific \\ Optimization};
    \node[minimum width=20mm, minimum height=22mm, dashed, draw=cornflowerblue, thick, align=center] (YR) at (2.5, 2.3)  {};
	\node[text width=1.6cm, align=center] at (2.5, 1.4) {\textbf{Detection}};    
	
	\node[draw=black, fill=mistyrose, minimum width=18mm, minimum height=8mm, text width=18mm, align=center, inner sep=0,outer sep=0] at (4.9, 2.9) {LLVM Backend \\ Modification};
    \node[draw=black, fill=mistyrose, minimum width=18mm, minimum height=8mm, text width=18mm, align=center, inner sep=0,outer sep=0] at (4.9, 2.0) {Register \\ Allocation};
    \node[minimum width=20mm, minimum height=22mm, dashed, draw=coralpink, thick, align=center] (SR) at (4.9, 2.3) {};
    \node[text width=1.6cm, align=center] at (4.9, 1.4) {\textbf{Mitigation}};

    \node[minimum width=46mm, minimum height=24mm, draw=black, thick, align=center] (BR) at (3.7, 2.3) {};

    \node[draw=black, fill=champagne, minimum width=16mm, minimum height=8mm, text width=16mm, align=center, inner sep=0,outer sep=0] (L0) at (7.2, 2.3) {Leakage-free \\ Assembly};       
    
   \draw [arrow] (U0.358) -- (K0.170); 
   \draw [arrow] (S0.east) -- (K1.west);
   \draw [thick] (D0.center) -- (D1.center);
   \draw [arrow] (D0.center) -- (YR.south);
   \draw [arrow] (D1.center) -- (SR.south);
   \draw [thick] (P1.344) -- (D0.center);
   \draw [arrow] (SR.0) -- (L0.west);
   \draw [arrow] (YR.east) -- (SR.west);

  
  

\end{tikzpicture}}
\caption{Overview of our secure compilation method}
\label{overview}
\vspace{-2ex}
\end{figure}
\setlength{\textfloatsep}{15pt}

Figure~\ref{overview} shows an overview of our method, which takes a
program $P$ as input and returns the mitigated code as output.  It has
two major steps.
First, sound type inference is used to detect leaks by assigning each
variable a \emph{distribution} type. User only provides an initial
annotation of input variables, i.e., \emph{public} (e.g.,
plaintext), \emph{secret} (e.g., key), or \emph{random} (e.g., mask),
while the types of other variables are inferred automatically.
Based on the inferred types, we check for each pair $(v_1,v_2)$ of
variables to see whether the values may be stored in the same register and
cause leaks.  If the pair is found to be leaky, we constrain the
compiler's backend register allocation modules to ensure that $v_1$
and $v_2$ are assigned to different registers or spilled to memory.

Our method differs from existing approaches in several aspects.
First, it specifically targets power side-channel leaks caused by
reuse of CPU registers in compilers, which have been largely
overlooked by prior work.
Second, it leverages Datalog, together with a number of
domain-specific optimizations, to achieve high efficiency and accuracy
during leak detection, where type inference rules are designed
specifically to capture register reuse related leaks.
Third, mitigation in the backend is systematic and leverages the
existing production-quality modules in LLVM to ensure that the
compiled code is secure by construction.

Unlike existing techniques that require \emph{a priori} translation of
the input program to a Boolean representation, our method works
directly on the word-level IR and thus fits naturally into modern
compilers.
For each program variable, the amount of leak is quantified using the
well-known Hamming Weight (HW) and Hamming Distance (HD) leakage
models~\cite{MangardOP07,mangard2002simple}. Correlation between these
models and leaks on real devices has been confirmed in prior works (see
Section~\ref{sec:motivation}).
We shall also show, via experiments, that leaks targeted by our method
exist even in program equipped with high-order
masking~\cite{barthe2015verified,barthe2016strong, balasch2014cost}.

To detect leaks quickly, we rely on type inference, which models the
input program using a set of Datalog facts and codifies the type
inference algorithm in a set of Datalog rules.  Then, an off-the-shelf
Datalog solver is used to deduce new facts.
Here, a domain-specific optimization, for example, is to leverage the
compiler's backend modules to extract a map from variables to
registers and utilize the map to reduce the computational overhead,
e.g., by checking pairs of some (instead of all) variables for leaks.

Our mitigation in the compiler's backend is systematic: it ensures
that all leaks detected by type inference are eliminated.  This is
accomplished by constraining register allocation modules and then
propagating the effect to subsequent modules, without having to
implement any new backend module from scratch.  Our mitigation is also
efficient in that we add a number of optimizations to ensure that the
mitigated code is compact and has low runtime overhead.
While our implementation focuses on x86, the technique itself is
general enough that it may be applied to other instruction set
architectures (ISAs) such as ARM and MIPS as well.

We have evaluated our method on a set of cryptographic software
benchmarks~\cite{barthe2015verified,bayrak2013sleuth}, including
implementations of well-known ciphers such as AES and MAC-Keccak.
These benchmark programs are all protected by masking countermeasures
but, still, we detected register reuse related leaks in the LLVM
compiled code.  The code produced by our mitigation, also based on
LLVM, is always leak free.
In terms of performance, our method significantly outperformed
competing approaches such as high-order masking in that our mitigated
code not only is more compact and secure, but also runs significantly
faster than code mitigated by high-order masking
techniques~\cite{barthe2015verified,barthe2016strong}.

To summarize, we make the following contributions:
\begin{itemize}
\item
We show that register reuse introduces side-channel leaks even in
software already protected by masking.
\item 
We propose a Datalog based type inference system to soundly and
quickly detect these side-channel leaks.
\item 
We propose a mitigation technique for the compiler's backend modules
to systematically remove the leaks. 
\item 
We implement the method in LLVM and show its effectiveness on a set of
cryptographic software.
\end{itemize}

The remainder of the paper is organized as follows.  First, we
illustrate the problem and the technical challenges associated with
solving it in Section~\ref{sec:motivation}.  Then, we review the
background including the threat model and leakage model in
Section~\ref{sec:preliminaries}.  Next, we present our method for leak
detection in Section~\ref{sec:detection} and leak mitigation in
Section~\ref{sec:mitigation}, followed by domain-specific
optimizations in Section~\ref{sec:optimization}.  We present our
experimental results in Section~\ref{sec:experiments}, review the
related work in Section~\ref{sec:related}, and give our conclusions in
Section~\ref{sec:conclusions}.

\section{Motivation}
\label{sec:motivation}

We use examples to illustrate why \emph{register reuse} may lead to
side-channel leaks and the challenges for removing them.

\subsection{The HW and HD Leaks}

Consider the program \emph{Xor()} in
Figure~\ref{fig:motivatingExample}, which takes the public
\emph{txt} and the secret \emph{key}  as input and returns the
exclusive-or of them as output.
Since logical 1 and 0 bits in a CMOS circuit correspond to different
leakage currents, they affect the power consumption of the
device~\cite{mangard2002simple}; such leaks were confirmed by prior
works~\cite{moradi2014side,brier2004correlation} and summarized in the
Hamming Weight (HW) model.
In program \emph{Xor()}, variable \emph{t} has a power side-channel
leak because its register value depends on the secret \emph{key}.

\begin{figure}
\vspace{1ex}
\centering
\begin{subfigure}{0.47\textwidth}
\begin{lstlisting}[basicstyle=\ttfamily\scriptsize]
//'txt': PUBLIC, 'key': SECRET and 't' is HW-sensitive 
uint32 Xor(uint32 txt, uint32 key) {uint32 t = txt ^ key; return t;}
//random variable 'mask1' splits 'key' to secure shares {mask1,mk}
uint64 SecXor(uint32 txt, uint32 key, uint32 mask1) {
   uint32 mk = mask1 ^ key;  // mask1^key
   uint32 t = txt ^ mk;      // txt^(mask1^key)
   return (mask1,t);           
}

//'mask1' splits 'key' to shares {mask1,mk} a priori 
//'mask2' splits the result to shares {mask2,t3} before return
uint64 SecXor2(uint32 txt, uint32 mk, unit32 mask1, unit32 mask2) {
   uint32 t1 = txt ^ mk; // txt^(mask1^key)
   uint32 t2 = t1 ^ mask2; // (txt^mask1^key)^mask2
   unit32 t3 = t2 ^ mask1; // (txt^mask1^key^mask2)^mask1
   return {mask2,t3};     
}
\end{lstlisting}
\end{subfigure}

\begin{subfigure}{0.48\textwidth}
\resizebox{\textwidth}{!}{
\begin{tabular}{c||c|c|c}
\hline
Name & Approach & HW-Sensitive & HD-Sensitive \\
\hline
Xor & No Masking & \cmark & \cmark \\
\hline
SecXor & First Order Masking & \xmark & \cmark \\
\hline
SecXor2 & Specialized Hardware \& Masking & \xmark & \cmark \\
\hline
\end{tabular}
}
\end{subfigure}
\vspace{-2ex}
\caption{Implementations of an XOR computation in the presence of HW and HD power side-channel leaks.}
\label{fig:motivatingExample}
\vspace{-1em}
\end{figure}

The leak may be mitigated
by \emph{masking}~\cite{goubin2001sound,akkar2003generic} as shown in
program \emph{SecXor()}.  The idea is to split a secret to $n$
randomized shares before using them; unless the attacker has all $n$
shares, it is theoretically impossible to deduce the secret.
In \emph{first-order} masking, the secret \emph{key} may be split
to \emph{\{mask1,mk\}} where \emph{mask1} is a random
variable, \emph{mk=mask1$\oplus$key} is the bit-wise Exclusive-OR
of \emph{mask1} and \emph{key}, and thus \emph{mask1$\oplus$mk=key}.
We say that \emph{mk} is \emph{masked} and thus \emph{leak free}
because it is statistically independent of the value of
the \emph{key}: if \emph{mask1} has a uniform random distribution then
so is \emph{mk}.  Therefore, when \emph{mk} is aggregated over time,
as in side-channel attacks, the result reveals no information
of \emph{key}.

Unfortunately, there can be leaks in \emph{SecXor()} when the
variables share a register and thus create second-order correlation.
For example, the x86 assembly code of \emph{mk$=$mask1$\oplus$key}
is \texttt{MOV~mask1~\%edx; XOR~key~\%edx}, meaning the values stored
in \texttt{\%edx} are \emph{mask1} and \emph{mask1$\oplus$key},
respectively.
Since bit-flips in the register also affect the leakage current, they
lead to side-channel leaks.  This is captured by the Hamming Distance
(HD) power model~\cite{brier2004correlation}:
\emph{HD(mask1,mask1$\oplus$key)} $=$ \emph{HW(mask1 
$\oplus$ (mask1 $\oplus$ key))} $=$ \emph{HW(key)},
which reveals \emph{key}.
Consider, for example, where \emph{key} is ${0001}_{b}$ and \emph{mask1} is ${1111}_b$ in binary.
If a register stores \emph{mask1} (=$1111_b$) first and updates its value as \emph{mask1$\oplus$key} (=$1110_b$), the transition of the register (bit-flip) is $0001_b$, which is same as the \emph{key} value.
%

In embedded systems, specialized
hardware~\cite{RuhrmairBK10,MaitiS11,Anagnostopoulos18} such as
physically unclonable function (PUF) and true random number generator
(TRNG) may produce \emph{key} and \emph{mask1} and map them to the
memory address space; thus, these variables are considered leak free.
Specialized hardware may also directly produce the masked
shares \emph{\{mask1,mk\}} without producing the unmasked
\emph{key} in the first place.  This more secure approach is  shown in
program \emph{SecXor2()}, where masked shares are used to compute the
result (\emph{txt$\oplus$key}), which is also masked, but by
\emph{mask2} instead of \emph{mask1}.

Inside \emph{SecXor2()}, care has been given to randomize the
intermediate results by \emph{mask2} first, before de-randomize them
by \emph{mask1}.  Thus, the CPU's registers never hold any unmasked
result. 
However, there can still be HD leaks, for example, when the same
register holds the following pairs at consecutive time steps:
\emph{(mask1,mk)}, \emph{(mask1,t1)}, and \emph{(mask2,t3)}.

\subsection{Identifying the HD Leaks}

To identify these leaks, we need to develop a scalable method.
While there are techniques for detecting flaws in various masking
implementations~\cite{coron2013higher,
hou2017improved,barthe2016strong,barthe2017parallel,duc2015making,
bloem2018formal,goubin2001sound,blomer2004provably,schramm2006higher,
canright2008very,rivain2010provably,prouff2013masking,
eldib2014synthesis,reparaz2015consolidating}, none of them were
scalable enough for use in real compilers, and none of them targeted
the HD leaks caused by register reuse.  Our work bridges the gap.

First, we check if there are sensitive, unmasked values stored in a
CPU's register.  Here, \emph{mask} means that a value is made
statistically independent of the secret using randomization.  We say
that a value is \emph{HW-sensitive} if, statistically, it still
depends on the secret.  For example, in
Figure~\ref{fig:motivatingExample}, \emph{key} is HW-sensitive
whereas \emph{mk=mask1$\oplus$key} has been masked.  If there
were \emph{nk=mask1$\vee$key}, it would be called HW-sensitive because
the masking is not perfect.

Second, we check if there is any pair of values $(v_1,v_2)$ that, when
stored in the same register, may cause an HD leak. That is,
$HD(v_1,v_2) = HW(v_1\oplus v_2)$ may statistically depend on the
secret.  For example, in Figure~\ref{fig:motivatingExample}, \emph{mk}
and \emph{mask1} form a HD-sensitive pair.

\paragraph{Formal Verification}

In general, deciding whether a variable is HW-sensitive, or a pair of
variables is HD-sensitive, is NP-hard, since it corresponds
to \emph{model counting}~\cite{zhang2018scinfer,eldib2014formal}.
This is illustrated by Table~\ref{tbl:truth}, which shows the truth
table of Boolean functions \emph{t1}, \emph{t2} and \emph{t3} in terms
of secret variable \emph{k} and random variables \emph{m1}, \emph{m2}
and \emph{m3}.
First, there is no HW leak because, regardless of whether \emph{k=}0
or 1, there is a 50\% chance of \emph{t1} and \emph{t2} being 1 and a
25\% chance of \emph{t3} being 1. This can be confirmed by counting 
the number of 1's in the top and bottom halves of the table.


When two values $(t1,t2)$ are stored in the same register, however,
the bit-flip may depend on the secret.  As shown in the column
$HD(t1,t2)$ of the table, when $k=0$, the bit is never flipped;
whereas when $k=1$, the bit is always flipped.
The existence of HD leak for $(t1,t2)$ can be decided by model
counting over the function $f_{t1\oplus t2}(k,m1,m2,m3)$: the number
of solutions is 0/8 for $k=0$ but 8/8 for $k=1$.
In contrast, there is no HD leak for $(t2,t3)$ because the number of
satisfying assignments (solutions) is always 2/8 regardless of
whether $k=0$ or $k=1$.

\paragraph{Type Inference}

Since model counting is expensive, e.g., taking hours or longer even
for small programs, it is not suitable for a compiler.  Thus, we
develop a fast, sound, and static type inference system to identify
the HD-sensitive pairs in a program.  By fast, we mean that our method
relies on syntactic information of the program or the platform (e.g.,
mapping from variables to physical registers).  By sound, we mean that
our method is conservative: it may introduce false alarms, and thus
may mitigate unnecessarily, but it never misses real leaks.

Specifically, we assign each variable one of three types: $\RUD$,
$\SID$ or $\UKD$ (details in Section~\ref{sec:preliminaries}).
Briefly, $\RUD$ means random uniform distribution, $\SID$ means secret
independent distribution, and $\UKD$ means unknown distribution.
Therefore, a variable may have a leak only if it is the $\UKD$ type.

\begin{table}[t]
\vspace{1ex}
\caption{Truth table showing that (1) there is no HW leak in \emph{t1,t2,t3} but (2) there is an HD leak when 
\vspace{-2ex}
\emph{t1,t2} share a register.}
\label{tbl:truth}
\centering
\scalebox{0.65}{
\begin{tabular}{|c|c|c|c||c|c|c||c|c|}
\hline
k & m1 & m2 & m3 & t1=             & t2=                    & t3=             &    HD(t1,t2)  &    HD(t2,t3)  \\              
        &    &    &    &    m1$\oplus$m2 & \ \ t1$\oplus$k \ \ &    t2$\wedge$m3 & =t1$\oplus$t2 & =t2$\oplus$t3 \\ \hline\hline 

0 & 0 & 0 & 0 & 0 & 0 & 0 & \textbf{0} & 0 \\ \hline 
0 & 0 & 0 & 1 & 0 & 0 & 0 & \textbf{0} & 0 \\ \hline 
0 & 0 & 1 & 0 & 1 & 1 & 0 & \textbf{0} & 1 \\ \hline 
0 & 0 & 1 & 1 & 1 & 1 & 1 & \textbf{0} & 0 \\ \hline 
0 & 1 & 0 & 0 & 1 & 1 & 0 & \textbf{0} & 1 \\ \hline 
0 & 1 & 0 & 1 & 1 & 1 & 1 & \textbf{0} & 0 \\ \hline 
0 & 1 & 1 & 0 & 0 & 0 & 0 & \textbf{0} & 0 \\ \hline 
0 & 1 & 1 & 1 & 0 & 0 & 0 & \textbf{0} & 0 \\ \hline
\hline 
1 & 0 & 0 & 0 & 0 & 1 & 0 & \textbf{1} & 1 \\ \hline 
1 & 0 & 0 & 1 & 0 & 1 & 1 & \textbf{1} & 0 \\ \hline 
1 & 0 & 1 & 0 & 1 & 0 & 0 & \textbf{1} & 0 \\ \hline 
1 & 0 & 1 & 1 & 1 & 0 & 0 & \textbf{1} & 0 \\ \hline 
1 & 1 & 0 & 0 & 1 & 0 & 0 & \textbf{1} & 0 \\ \hline 
1 & 1 & 0 & 1 & 1 & 0 & 0 & \textbf{1} & 0 \\ \hline 
1 & 1 & 1 & 0 & 0 & 1 & 0 & \textbf{1} & 1 \\ \hline 
1 & 1 & 1 & 1 & 0 & 1 & 1 & \textbf{1} & 0 \\ \hline
\hline
\ \ $\UKD$\ \  & \ \ $\RUD$\ \  & \ \ $\RUD$\ \  & \ \ $\RUD$\ \  & $\RUD$ & $\RUD$ & $\SID$ & $\UKD$ & $\SID$* \\\hline
\end{tabular}
}

\vspace{1ex}
\raggedright \scriptsize{* Our Datalog based type inference rules can infer it as $\SID$ instead of $\UKD$}

\end{table}

In Table~\ref{tbl:truth}, for example, given $t1 \leftarrow m1 \oplus
m2$, where $m1$ and $m2$ are random ($\RUD$), it is easy to see that
$t1$ is also random ($\RUD$).
For $t3 \leftarrow t2 \wedge m3$, where $t2$, $m3$ are $\RUD$, however,
$t3$ may not always be random, but we can still prove that $t3$ is
$\SID$; that is, $t3$ is statistically independent of $k$.
This type of \emph{syntactical} inference is fast because it does not
rely on any \emph{semantic} information, although in general, it is
not as accurate as the model counting based approach.
Nevertheless, such inaccuracy does not affect the soundness of our
mitigation.

Furthermore, we rely on a Datalog based declarative analysis
framework~\cite{whaley2005using,zhang2014abstraction,lam2005context,whaley2004cloning,bravenboer2009strictly}
to implement and refine the type inference rules, which can infer
$HD(t2,t3)$ as $\SID$ instead of $\UKD$.  We also leverage
domain-specific optimizations, such as precomputing certain Datalog
facts and using compiler's backend information, to reduce cost and
improve accuracy.
%

\vspace{-1mm}

\subsection{Mitigating the HD Leaks}

\begin{figure}
\centering
\begin{subfigure}{.21\textwidth}
\centering
\begin{lstlisting}[escapechar=!, numbers=left,frame=tlrb, numberstyle=\tiny, basicstyle=\ttfamily\scriptsize]{Name}		
	// assembly for Table1 		
	movl	%edi, -4(%rbp)
	movl	%esi, -8(%rbp)
	movl	%edx, -12(%rbp)
	movl	%ecx, -16(%rbp)
	movl	-4(%rbp), %eax
	xorl	-8(%rbp), %eax
	movl	%eax, -20(%rbp)
	!\colorbox{pastelorange}{xorl  -16(\%rbp), \%eax }!
	movl	%eax, -24(%rbp)
	andl	-12(%rbp), %eax
	movl	%eax, -28(%rbp)
	
	popq	%rbp
\end{lstlisting}
\vspace{-1ex}
\caption{Before Mitigation}
\label{fig:assemblyCompare-before}
\end{subfigure}
\hspace{7mm}
\begin{subfigure}{0.21\textwidth}
\centering
\begin{lstlisting}[firstnumber=1, escapechar=!, numbers=left, frame=tlrb, numberstyle=\tiny, basicstyle=\ttfamily\scriptsize]{Name}
	// assembly for Table1			
	movl	%edi, -4(%rbp)
	movl	%esi, -8(%rbp)
	movl	%edx, -12(%rbp)
	movl	%ecx, -16(%rbp)
	movl	-4(%rbp), %eax
	xorl	-8(%rbp), %eax
	movl	%eax, -20(%rbp)
	!\colorbox{pastelorange}{xorl	\%eax, -16(\%rbp)}!
	movl -16(%rbp), %ecx
	andl	-12(%rbp), %ecx
	movl	%ecx, -28(%rbp)
	movl -28(%rbp), %eax
	popq	%rbp
\end{lstlisting}
\vspace{-1ex}
\caption{After Mitigation}
\label{fig:assemblyCompare-after}
\end{subfigure}

\vspace{3mm}

\begin{subfigure}{0.48\textwidth}
\centering
\resizebox{\textwidth}{!}{
\tikzstyle{arrow}=[thick,->,>=stealth,black]
\begin{tikzpicture}[font=\scriptsize, inner sep=0, outer sep=0]

\node[text width=15mm, align=center] at (0.4, -0.3) {stack};
\node[draw=black, minimum width=12mm, minimum height=2mm, text width=12mm, align=center, inner sep=0,outer sep=0]  at (0.4, -0.65) {...};  
\node[draw=black, minimum width=12mm, minimum height=5mm, text width=12mm, align=center, inner sep=0,outer sep=0]  at (0.4, -1.0) {m1};  
\node[draw=black, minimum width=12mm, minimum height=5mm, text width=12mm, align=center, inner sep=0,outer sep=0]  at (0.4, -1.5) {m2};  
\node[draw=black, minimum width=12mm, minimum height=5mm, text width=12mm, align=center, inner sep=0,outer sep=0]  at (0.4, -2.0) {m3};  
\node[draw=black, minimum width=12mm, minimum height=5mm, text width=12mm, align=center, inner sep=0,outer sep=0]  at (0.4, -2.5) {key};  
\node[draw=black, minimum width=12mm, minimum height=5mm, text width=12mm, align=center, inner sep=0,outer sep=0]  at (0.4, -3.0) {m1$\oplus$m2};
\node[draw=black, minimum width=12mm, minimum height=2mm, text width=12mm, align=center, inner sep=0,outer sep=0]  at (0.4, -3.35) {...};  

\node[align=center,text width=8mm, align=left, inner sep=0, outer sep=0, font=\tiny] at (1.6, -1.0) {-4(\%rbp)};
\node[align=center,text width=8mm, align=left, inner sep=0, outer sep=0, font=\tiny] at (1.6, -1.5) {-8(\%rbp)};
\node[align=center,text width=8mm, align=left, inner sep=0, outer sep=0, font=\tiny] at (1.6, -2.0) {-12(\%rbp)};
\node[align=center,text width=8mm, align=left, inner sep=0, outer sep=0, font=\tiny] at (1.6, -2.5) {-16(\%rbp)};
\node[align=center,text width=8mm, align=left, inner sep=0, outer sep=0, font=\tiny] at (1.6, -3.0) {-20(\%rbp)};

\node[text width=15mm, align=center] at (-1.1, -1.5) {\%eax};
\node[draw=black, minimum width=12mm, minimum height=5mm, text width=12mm, align=center, inner sep=0,outer sep=0, fill=orange!50]  at (-1.1, -2.0) {m1$\oplus$m2};  

\node[text width=30mm, align=center, inner sep=0, outer sep=0] at (0.2, -3.8) {\textbf{After executing line 8}};

\node[draw=black, dashed, minimum width=40mm, minimum height=40mm, inner sep=0, outer sep=0] (R0) at (0.15, -2.0) {};


\node[text width=15mm, align=center] at (7.5, 0.7) {stack};
\node[draw=black, minimum width=15mm, minimum height=2mm, text width=15mm, align=center, inner sep=0,outer sep=0]  at (7.5, 0.35) {...};  
\node[draw=black, minimum width=15mm, minimum height=5mm, text width=15mm, align=center, inner sep=0,outer sep=0]  at (7.5, 0.0) {key};  
\node[draw=black, minimum width=15mm, minimum height=5mm, text width=15mm, align=center, inner sep=0,outer sep=0]  at (7.5, -0.5) {m1$\oplus$m2};
\node[draw=black, minimum width=15mm, minimum height=2mm, text width=15mm, align=center, inner sep=0,outer sep=0]  at (7.5, -0.85) {...};  


\node[align=center, font=\tiny] at (8.8, -0.0) {-16(\%rbp)};
\node[align=center, font=\tiny] at (8.8, -0.5) {-20(\%rbp)};

\node[text width=15mm, align=center] at (5.7, 0.25) {\%eax};
\node[draw=black, minimum width=15mm, minimum height=5mm, text width=15mm, align=center, inner sep=0,outer sep=0, fill=orange!50]  at (5.7, -0.25) {m1$\oplus$m2$\oplus$key}; 

\node[text width=40mm, align=center] at (7.0, -1.4) {\textbf{Before Mitigation} \\ (after executing line 9)};

\node[draw=black, dashed, minimum width=48mm, minimum height=27mm] (R1) at (7.0, -0.45) {};


\node[text width=15mm, align=center] at (7.5, -2.5) {stack};
\node[draw=black, minimum width=15mm, minimum height=2mm, text width=15mm, align=center, inner sep=0,outer sep=0]  at (7.5, -2.85) {...};  
\node[draw=black, minimum width=15mm, minimum height=5mm, text width=15mm, align=center, inner sep=0,outer sep=0]  at (7.5, -3.2) {m1$\oplus$m2$\oplus$key};  
\node[draw=black, minimum width=15mm, minimum height=5mm, text width=15mm, align=center, inner sep=0,outer sep=0]  at (7.5, -3.7) {m1$\oplus$m2};
\node[draw=black, minimum width=15mm, minimum height=2mm, text width=15mm, align=center, inner sep=0,outer sep=0]  at (7.5, -4.05) {...};  


\node[align=center, font=\tiny] at (8.8, -3.2) {-16(\%rbp)};
\node[align=center, font=\tiny] at (8.8, -3.7) {-20(\%rbp)};

\node[text width=15mm, align=center] at (5.7, -2.95) {\%eax};
\node[draw=black, minimum width=15mm, minimum height=5mm, text width=15mm, align=center, inner sep=0,outer sep=0, fill=orange!50]  at (5.7, -3.45) {m1$\oplus$m2}; 

\node[text width=40mm, align=center] at (7.0, -4.6) {\textbf{After Mitigation} \\ (after executing line 9)};

\node[draw=black, dashed, minimum width=48mm, minimum height=27mm] (R2) at (7.0, -3.65) {};

\draw[thick,->,>=stealth,red] (R0.east) -- (R1.west) node[midway,sloped,above, yshift=1mm, red] {\textbf{HD = key (leak)}};
\draw[thick,->,>=stealth,black] (R0.east) -- (R2.west) node[midway,sloped,above, yshift=1mm] {HD = 0};

\end{tikzpicture}
}%
\caption{Diagram for stack and register \%eax}
\label{fig:assemblyCompare-dia}
\end{subfigure}
\vspace{-2ex}
\caption{The assembly code before and after mitigation.}
\label{fig:assemblyCompare}
\end{figure}
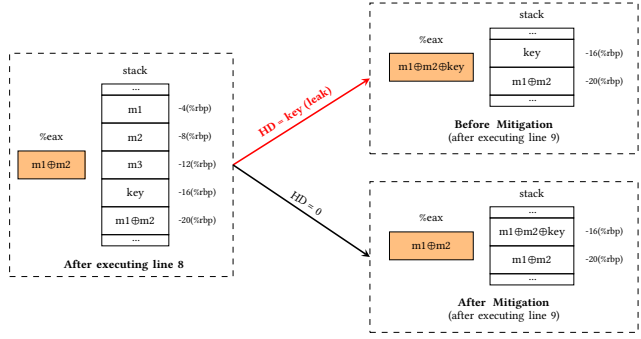

To remove the leaks, we constrain the register allocation algorithm
using our inferred types.  We focus on LLVM and x86, but the method is
applicable to MIPS and ARM as well.
To confirm this, we inspected the assembly code produced by LLVM for
the example (\emph{t1,t2,t3}) in Table~\ref{tbl:truth} and found HD
leaks on all three architectures.
For x86, in particular, the assembly code is shown in
Figure~\ref{fig:assemblyCompare-before}, which uses \%eax to store all
intermediate variables and thus has a leak in \emph{HD(t1,t2)}.

\setlength{\textfloatsep}{5pt}

Figure~\ref{fig:assemblyCompare-after} shows our mitigated code, where
the HD-sensitive variables \emph{t1} and \emph{t2} are stored in
different registers.  Here, \emph{t1} resides in
\%eax and memory -20(\%rbp) whereas \emph{t2} resides
in \%ecx and memory -16(\%rbp).
The stack and a value of \%eax are shown in
Figure~\ref{fig:assemblyCompare-dia}, both before and after
mitigation, when the leak may occur at lines 8-9. 
%
Since the value of \emph{k} is used only once in the example, i.e.,
for computing \emph{t2}, overwriting its value stored in the original
memory location -16(\%rbp) does not affect subsequent execution.
If \emph{k} were to be used later, our method would have made a copy
in memory and direct uses of \emph{k} to that memory location.

Register allocation in real compilers is a highly optimized process.
Thus, care must be given to maintain correctness and performance.
For example, the naive approach of assigning all HD-sensitive
variables to different registers does not work because the number of
registers is small (x86 has 4 general-purpose registers while MIPS has
24) while the number of sensitive variables is often large, meaning
many variables must be \emph{spilled} to memory.

The instruction set architecture also add constraints.  In x86, for
example, \%eax is related to \%ah and \%al and thus cannot be assigned
independently. Furthermore, binary operations such as \emph{Xor} may
require that the result and one operand must share the same register
or memory location. Therefore, for \emph{mk=mask1$\oplus$key}, it
means that either \emph{mk} and \emph{mask1} share a register, which
causes a leak in \emph{HD(mk, mask1)=HW(key)}, or \emph{mk}
and \emph{key} share a register, which causes a leak in \emph{HW(key)}
itself.
%
Thus, while modifying the backend, multiple submodules must be
constrained together to ensure the desired register and memory
isolation (see Section~\ref{sec:mitigation}).

\subsection{Leaks in High-order Masking}

Here, a question is whether the HD leak can be handled by second-order
masking (which involves two variables).  The answer is no, because
even with high-order masking techniques such as Barthe et
al.~\cite{barthe2015verified, barthe2016strong,barthe2017parallel},
the compiled code may still have HD leaks introduced by register
reuse.  We confirmed this through experiments, where the code compiled
by LLVM for high-order masked programs from Barthe et
al.~\cite{barthe2015verified} was found to contain HD leaks.


Figure~\ref{secondOrder} illustrates this problem on a second-order
arithmetic masking of the multiplication of \texttt{txt} (public)
and \texttt{key} (secret) in a finite field.  Here, the symbol $\ast$
denotes multiplication.
While there are a lot of details, at a high level, the program relies
on the same idea of \emph{secret sharing}: random variables are used
to split the secret \emph{key} to three shares, before these shares
participate in the computation.  The result is a masked
triplet \emph{(res0,res1,res2)} such
that \emph{(res0$\oplus$res1$\oplus$res2)$=$key$*$txt}.

The x86 assembly code in Figure~\ref{secondOrder} has leaks because
the same register \%edx stores both \emph{mask0 $\oplus$ mask1}
and \emph{mask0 $\oplus$ mask1 $\oplus$ key}.  Let the two values be
denoted \%edx$_1$ and \%edx$_2$, we have \emph{HD(\%edx$_1$,\%edx$_2$)
= HW(key)}.
%
%
Similar leaks exist in the LLVM-generated assembly code of this
program for ARM and MIPS as well, but we omit them for brevity.

\begin{figure}
 
\begin{minipage}{.46\textwidth}
\begin{lstlisting}[escapechar=!, numbers=left, frame=tlrb, numberstyle=\tiny, basicstyle=\ttfamily\scriptsize]
uint8 SecondOrderMaskingMultiply(uint8 txt, uint8 key) {
	int mask0, mask1, mask2, mask3, mask4, mask5, mask6; //random
	int t1 =  mask0 ^ mask1 ^ key;
	int t2 =  mask2 ^ mask3 ^ txt;
	int t3 = (mask4 ^ mask0 * mask3) ^ mask1 * mask2;
	int t4 = (mask5 ^ mask0 * t2) ^ t1 * mask2;
	int t5 = (mask6 ^ mask1 * t2) ^ t1 * mask3;
	res0 = (mask0 * mask2 ^ mask4) ^ mask5;
	res1 = (mask1 * mask3 ^ t3) ^ mask6;
	res2 = (t1 * t2 ^ t4) ^ t5;
	return {res0, res1, res2};
}
\end{lstlisting}
\end{minipage}
 
\begin{minipage}{.46\textwidth}
\begin{lstlisting}[escapechar=!, basicstyle=\ttfamily\scriptsize]
	movzbl	-41(%rbp), %edx  // mask0 is loaded to %edx
	movzbl	-43(%rbp), %esi   // mask1 is loaded to %esi
	!\colorbox{pastelorange}{xorl	\%esi, \%edx}!			// mask0^mask1 is stored to %edx !(\%edx$_1$)!
	movzbl	-44(%rbp), %esi	// key is loaded to %esi
	!\colorbox{pastelorange}{xorl	\%esi, \%edx}!			// mask0^mask1^key is stored to %edx !(\%edx$_2$)!	
	movb	%dl, %al
	movb	%al, -50(%rbp)
\end{lstlisting}
\end{minipage}
\vspace{-2ex}
\caption{Second-order masking of multiplication in a finite field, and the LLVM-generated x86 assembly code of Line 3.}
\label{secondOrder}
\end{figure}

\ignore{

\begin{figure}
\vspace{1ex}   
\begin{lstlisting}[numbers=left,numberstyle=\tiny, basicstyle=\ttfamily\scriptsize]
	ldrb	r1, [sp, #31]			//reg r1, [sp, #31] store mask0	
	ldrb	r0, [sp, #29]			// reg r0, [sp, #29] store mask1
	eor	r0, r1, r0				// reg r0 stores mask0 ^ mask1 
	ldrb	r1, [sp, #28]			// reg r1, [sp, #28] store key
	eor	r0, r0, r1				// reg r0 stores mask0 ^ mask1 ^ key
	strb	r0, [sp, #22]
\end{lstlisting}
\caption{Second Order Masking Assembly (ARM)}
\label{secondOrderAssemblyARM}
\end{figure}

}

\section{Preliminaries}
\label{sec:preliminaries}

We define the threat model and then review the leakage models used for
quantifying the power side channel.

\subsection{The Threat Model}
\label{sec:threatmodel}

We assume the attacker has access to the software code, but not the
secret data, and the attacker's goal is to gain information of the
secret data.  The attacker may measure the power consumption of a
device that executes the software, at the granularity of each machine
instruction.  A set of measurement traces is aggregated to perform
statistical analysis, e.g., as in DPA attacks.
In mitigation, our goal is to eliminate the statistical dependence
between secret data and the (aggregated) measurement data.

Let $P$ be the program under attack and the triplet
$(\mathbf{x}, \mathbf{k}, \mathbf{r})$ be the input: sets
$\mathbf{x}$, $\mathbf{k}$ and $\mathbf{r}$ consist
of \emph{public}, \emph{secret}, and \emph{random (mask)} variables,
respectively.
Let $x$, $k_1$, $k_2$, and $r$ be valuations of these input variables.
Then, ${\sigma}_{t} (P,x,k_1,r)$ denotes, at time step $t$, the power
consumption of a device executing $P$ under input $x$, $k_1$
and $r$.  Similarly, ${\sigma}_{t} (P,x,k_2,r)$ denotes the power
consumption of the device executing $P$ under input $x$, $k_2$ and
$r$. 
Between steps $t$ and $t+1$, one instruction in $P$ is executed.

We say $P$ has a leak if there are $t$, $x$, $k_1$ and $k_2$ such that
the distribution of ${\sigma}_{t} (P,x,k_1,r)$ differs from that of
${\sigma}_{t} (P,x,k_2,r)$.
Let random variables in $\mathbf{r}$ be uniformly distributed in the
domain $R$, and let the probability of each $r\in R$ be $Pr(r)$, we
expect
$\forall t, x, k_1, k_2 ~.~$

\vspace{-1em}
\begin{equation}
\sum_{ r\in R} { { \sigma  }_{ t } } (P, x, k_1, r) ~Pr(r) = \\
\sum_{ r\in R} { { \sigma  }_{ t } } (P, x, k_2, r) ~Pr(r)
\label{eq:1}
\end{equation}


For efficiency reasons, in this work, we identify \emph{sufficient
conditions} under which Formula~\ref{eq:1} is implied.  Toward this
end, we focus on the leaks of individual variables, and pairs of
variables, in $P$ instead of the sum $\sigma_t$: if we remove all
individual leaks, the leak-free property over the sum
$\sigma_t(P,x,k,r)$ is implied.

\subsection{The Leakage Model}

In the Hamming Weight (HW) model~\cite{MangardOP07,mangard2002simple},
the leakage associated with a register value, which corresponds to an
intermediate variable in the program, depends on the number of 1-bits.
Let the value be $D = \sum _{ i=0 }^{ n-1 }{ { d }_{ i }{ 2 }^{ i } }$
where $d_0$ is the least significant bit, $d_{n-1}$ is the most
significant bit, and each bit $ { d }_{ i }$, where $0 \leq i < n$, is
either 0 or 1.
The Hamming Weight of $D$ is $HW(D) = \sum _{ i=0 }^{ n-1 }{ { d }_{ i
} }$.

In the Hamming Distance (HD) model~\cite{MangardOP07,mangard2002simple}, the leakage depends not only on
the current register value $D$ but also a reference value $D'$.
Let $D' = \sum _{ i=0 }^{ n-1 }{ d_{ i }' { 2 }^{ i } }$.
We define the Hamming Distance between $D$ and $D'$ as $HD(D,D')
= \sum _{ i=0 }^{ n-1 }{ { d }_{ i } \oplus {d}_{i}' }$, which is
equal to $HW(D \oplus D')$, the Hamming Weight of the bit-wise XOR of
$D$ and $D'$.
Another interpretation is to regard $HW(D)$ as a special case of
$HD(D,D')$, where all bits in the reference value $D'$ are set to 0.

The widely used HW/HD models have been confirmed on various
devices~\cite{kocher1999differential,clavier2000differential,
brier2004correlation, messerges2000using,moradi2014side}.
The correlation between power variance and number of 1-bits may be
explained using the \emph{leakage current} of a CMOS transistor, which
is the foundation of modern computing devices.
Broadly speaking, a CMOS transistor has two kinds of leakage
currents: \emph{static} and \emph{dynamic}.  Static leakage current
exists all the time but the volume depends on whether the transistor
is on or off, i.e., a logical 1.  Dynamic leakage current occurs only
when a transistor is switched (0-1 or 1-0 flip).  While static leakage
current is captured by the HW model, dynamic leakage current is
captured by the HD model (for details refer to
Mangard~\cite{mangard2002simple}.)

\subsection{The Data Dependency}

We consider two dependency relations: \emph{syntactical}
and \emph{statistical}.
Syntactical dependency is defined over the program structure: a
 function $f(k,\ldots)$ syntactically depends on the variable $k$,
 denoted $\mathcal{D}_{\mathit{syn}}(f,k)$, if $k$ appears in the
 expression of $f$; that is, $k$ is in the support of $f$,
 denoted $k\in supp(f)$.
%
%
%

Statistical dependency is concerned with scenarios where random
variables are involved.  For example, when $f(k,r) = k \oplus r$, the
probability of $f$ being logical 1 (always 50\%) is not dependent on
$k$.  However, when $f(k,r) = k \vee r$, where $r$ is a random uniform
distribution in $[0,1]$, the probability of $f$ being logical 1 is
100\% when $k$ is 1, but 50\% when $k$ is 0. In the latter case, we
say that $f$ is statistically dependent on $k$, denoted
$\mathcal{D}_{\mathit{sta}}(f,k)$.

The relative strengths of the dependency relations are as follows:
%
%
$\neg \mathcal{D}_\mathit{syn}(f,k) \implies \neg \mathcal{D}_\mathit{sta}(f,k)$,
i.e., if $f$ is syntactically independent of $k$, it is statistically
independent of $k$.
%
%
%
%
In this work, we rely on $\mathcal{D}_{\mathit{syn}}$ to infer
$\mathcal{D}_{\mathit{sta}}$ during type inference, since the
detection of HD leaks must be both fast and sound.

\section{Type-based Static Leak Detection}
\label{sec:detection}

We use a type system that starts from the input annotation
($\mathit{IN_{PUBLIC}}$, $\mathit{IN_{SECRET}}$ and
$\mathit{IN_{RANDOM}}$) and computes a \emph{distribution type} for
all variables.  The type indicates whether a variable may
statistically depend on the secret input.

\subsection{The Type Hierarchy}

The distribution type of variable $v$, denoted $\mathsf{TYPE}(v)$, may
be one of the following kinds:
\begin{itemize}
\item
$\RUD$, which stands for \emph{random uniform distribution}, means $v$
is either a random input $m\in\mathit{IN_{RANDOM}}$ or perfectly
randomized~\cite{blomer2004provably} by $m$, e.g., $v
= \mathit{k} \oplus m$.
\item
$\SID$, which stands for \emph{secret independent distribution}, means
that, while not $\RUD$, $v$ is statistically independent of the secret
variable in $\mathit{IN_{\SECRET}}$.
\item
$\UKD$, which stands for \emph{unknown distribution}, indicates that
we are not able to prove that $v$ is $\RUD$ or $\SID$ and thus have to
assume that $v$ may have a leak.
\end{itemize}

The three types form a hierarchy:
$\UKD$ is the least desired because it means that a leak may exist.
$\SID$ is better: although it may not be $\RUD$, we can still prove
that it is statistically independent of the secret, i.e., no leak.
$\RUD$ is the most desired because the variable not only is
statistically independent of the secret (same as in $\SID$), but also
can be used like a random input, e.g., to mask other ($\UKD$)
variables.
For leak mitigation purposes, it is always sound to treat an $\RUD$
variable as $\SID$, or an $\SID$ variable as $\UKD$, although it may
force instructions to be unnecessarily mitigated.

In practice, we want to infer as many $\SID$ and $\RUD$ variables as
possible.
For example, if $k\in \mathit{IN_{\SECRET}}$, $m\in \mathit{IN_{RANDOM}}$ and $k_m = k \oplus m$,
then $\mathsf{TYPE}(k) = \UKD$ and $\mathsf{TYPE}(k_m) = \RUD$.
If $x\in\mathit{IN_{PUBLIC}}$ and $xk_m = x \wedge k_m$, then
$\mathsf{TYPE}(xk_m) = \SID$ because, although $x$ may have any
distribution, since $k_m$ is $\RUD$, $xk_m$ is statistically
independent of the secret.

We prefer $\RUD$ over $\SID$, when both are applicable to a variable
$x_1$, because if $x_1$ is XOR-ed with a $\UKD$ variable $x_2$, we can
easily prove that $x = x_1\oplus x_2$ is $\RUD$ using local inference,
as long as $x_1$ is $\RUD$ and $x_2$ is not randomized by the same
input variable.
However, if $x_1$ is labeled not as $\RUD$ but as $\SID$, local
inference rules may not be powerful enough to prove that $x$ is $\RUD$
or even $\SID$; as a result, we have to treat $x$ as $\UKD$ (leak),
which is less accurate.

\subsection{Datalog based Analysis}

In the remainder of this section, we present type inference for
individual variables first, and then for HD-sensitive pairs.

We use Datalog to implement the type inference. Here, program
information is captured by a set of relations called the \emph{facts},
which include the annotation of input in $\mathit{IN_{PUBLIC}}$
($\SID$), $\mathit{IN_{SECRET}}$ ($\UKD$) and $\mathit{IN_{RANDOM}}$
($\RUD$).
The inference algorithm is codified in a set of relations called
the \emph{rules}, which are steps for deducing types.  For example,
when $z=x\oplus m$ and $m$ is $\RUD$, $z$ is also $\RUD$ regardless of
the actual expression that defines $x$, as long as $m\not\in supp(x)$.
This can be expressed as an inference rule.

After generating both the facts and the rules, we combine them to form
a Datalog program, and solve it using an off-the-shelf Datalog engine.
Inside the engine, the rules are applied to the facts to generate new
facts (types); the iterative procedure continues until the set of
facts reaches a fixed point.

Since our type inference is performed on the LLVM IR, there are only a
few instruction types to consider.  For ease of presentation, we
assume that a variable $v$ is defined by either a unary operator or a
binary operator ($n$-ary operator may be handled similarly).
\begin{itemize}
\item  
$v \leftarrow \mathit{Uop}(v_1)$, where $\mathit{Uop}$ is a unary
operator such as the Boolean (or bit-wise) negation.
\item  
$v \leftarrow \mathit{Bop}(v_1,v_2)$, where $\mathit{Bop}$ is a binary
operator such as Boolean (or bit-wise) $\oplus$, $\wedge$, $\vee$ and
$\ast$ (finite-field multiplication).
\end{itemize}
For $v \leftarrow \mathit{Uop}(v_1)$, we have $\mathsf{TYPE}(v)
= \mathsf{TYPE}(v_1)$, meaning $v$ and $v_1$ have the same type.
For $v\leftarrow\mathit{Bop}(v_1,v_2)$, the type depends on
(1) if $\mathit{Bop}$ is $\mathit{Xor}$,
(2) if  $\mathsf{TYPE}(v_1)$ and $\mathsf{TYPE}(v_2)$ are $\SID$
or $\RUD$, and
(3) the sets of input variables upon which $v_1$ and $v_2$ depend.

\subsection{Basic Type Inference Rules}

Prior to defining the rules for $\mathit{Bop}$, we define two related
functions, $\mathsf{unq}$ and $\mathsf{dom}$, in addition to
$\mathsf{supp}(v)$, which is the set of input variables upon which $v$
depends syntactically.

\ignore{
\begin{defn}
\label{def:supp}
$\mathsf{supp}:V \rightarrow \mathit{IN}$ is a function that returns,
for each variable $v\in V$, the set of input variables that $v$ is
syntactically dependent on.  That is, if $v\in \mathit{IN}$,
$\mathsf{supp}(v) = \{v\}$;
\begin{itemize}
\item if $v\leftarrow \mathit{Uop}(v_1)$,  $\mathsf{supp}(v) = \mathsf{supp}(v_1)$; and
\item if $v\leftarrow \mathit{Bop}(v_1,v_2)$, $\mathsf{supp}(v) = \mathsf{supp}(v_1) \cup \mathsf{supp}(v_2)$.
\end{itemize}
\end{defn}
}

\begin{defn}
\label{def:unq}
$\mathsf{unq}:V \rightarrow \mathit{IN_{RANDOM}}$ is a function that
returns, for each variable $v\in V$, a subset of mask variables
defined as follows: if $v\in\mathit{IN_{RANDOM}}$, $\mathsf{unq}(v)
= \{v\}$; but if $v\in\mathit{IN}\setminus\mathit{IN_{RANDOM}}$,
$\mathsf{unq}(v) = \{~\}$;
\begin{itemize}
\item if $v\leftarrow \mathit{Uop}(v_1)$,  $\mathsf{unq}(v) = \mathsf{unq}(v_1)$; and
\item if $v\leftarrow \mathit{Bop}(v_1,v_2)$, $\mathsf{unq}(v) = \left(\mathsf{unq}(v_1) \cup \mathsf{unq}(v_2)\right) \setminus (\mathsf{supp}(v_1)$ $\cap$ $\mathsf{supp}(v_2) )$.
\end{itemize}
\end{defn}
\noindent
Given the data-flow graph of all instructions involved in computing
$v$ and an input variable $m\in\mathsf{unq}(v)$, there must exist a
unique path from $m$ to $v$ in the graph.  If there are more paths (or
no path), $m$ would not have appeared in $\mathsf{unq}(v)$.

\begin{defn}
\label{def:dom}
$\mathsf{dom}:V \rightarrow \mathit{IN_{RANDOM}}$ is a function that
returns, for each variable $v\in V$, a subset of mask variables
defined as follows:
if $v\in \mathit{IN_{RANDOM}}$, $\mathsf{dom}(v) = \{v\}$, but 
if $v\in \mathit{IN} \setminus \mathit{IN_{RANDOM}}$, then $\mathsf{dom}(v) = \{~\}$; 
\begin{itemize}
\item if $v \leftarrow \mathsf{Uop}(v_1)$,  $\mathsf{dom}(v) = \mathsf{dom}(v_1)$; and
\item if $v \leftarrow \mathsf{Bop}(v_1,v_2)$, where operator $\mathsf{Bop}=\mathsf{Xor}$, then $\mathsf{dom}(v) =$ $\left(\mathsf{dom}(v_1) \cup \mathsf{dom}(v_2)\right) \cap \mathsf{unq}(v)$; else $\mathsf{dom}(v) = \{~\}$. 
\end{itemize}
\end{defn}
\noindent
Given the data-flow graph of all instructions involved in computing
$v$ and an input variable $m \in \mathsf{dom}(v)$, there must exist a
unique path from $m$ to $v$, along which all binary operators are
$\mathsf{Xor}$; if there are more such paths (or no such path), $m$
would not have appeared in $\mathsf{dom}(v)$.

Following the definitions of $\mathsf{supp}$, $\mathsf{unq}$ and
$\mathsf{dom}$, it is straightforward to arrive at the basic inference
rules~\cite{OMHE17,barthe2016strong,zhang2018scinfer}:

\vspace{-3mm}
\centering
{\footnotesize
\[\begin{array}{l}
  {Rule}_{1} \myinferrule
                   {\mathsf{TYPE}(v) = \RUD} 
                   { \mathsf{dom}(v)\neq \emptyset }   \\
\\
\end{array}\]
\vspace{-3mm}
\[\begin{array}{l}
  {Rule}_{2} \myinferrule
                   {\mathsf{TYPE}(v) = \SID}
                   {\mathsf{supp}(v)\cap\mathit{IN_{\SECRET}} = \emptyset \wedge
                    \mathsf{TYPE}(v) \neq \RUD}
\end{array}\]
}
\justifying

Here, $\mathit{Rule_1}$ says if
$v=m\oplus \mathit{expr}$, where $m$ is a random input and
$\mathit{expr}$ is not masked by $m$, then $v$ has random uniform
distribution.  This is due to the property of XOR.
$\mathit{Rule_2}$ says if $v$ is syntactically independent of
variables in $\mathit{IN_{\SECRET}}$, it has a secret independent
distribution, provided that it is not $\RUD$.
%

\subsection{Inference Rules to Improve Accuracy}

With the two basic rules only, any variable not assigned $\RUD$ or
$\SID$ will be treated as $\UKD$, which is too conservative.
For example, $v = (k\oplus m) \wedge x$ where
$k\in\mathit{IN_{SECRET}}$, $m\in\mathit{IN_{RANDOM}}$ and
$x\in\mathit{IN_{PUBLIC}}$, is actually $\SID$.  This is because
$k\oplus m$ is random and the other component, $x$, is secret
independent.  Unfortunately, the two basic rules cannot infer that $v$
is $\SID$.  The following rules are added to solve this problem.

\centering
{\footnotesize
\[\begin{array}{l}

  {Rule}_{3a} \myinferrule
                   {\mathsf{TYPE}(v) = \SID}
                   {\begin{array}{c}
                    v\leftarrow \mathit{Bop}(v_1,v_2) \wedge  
                    \mathsf{supp}(v_1)\cap\mathsf{supp}(v_2) = \emptyset \\   
                    \wedge
                     \mathit{Bop} \not\in \{\mathit{Xor}, \mathit{GMul} \} \wedge
                    \mathsf{TYPE}(v_1)=\RUD \wedge
                    \mathsf{TYPE}(v_2)=\SID
                    \end{array}}\\\\

%
  {Rule}_{3b} \myinferrule
                   {\mathsf{TYPE}(v) = \SID}
                   {\begin{array}{c}
                    v\leftarrow \mathit{Bop}(v_1,v_2) \wedge  
                    \mathsf{supp}(v_1)\cap\mathsf{supp}(v_2) = \emptyset \\   
                    \wedge
                    \mathit{Bop} \not\in \{\mathit{Xor}, \mathit{GMul} \} \wedge
                      \mathsf{TYPE}(v_1)=\SID \wedge \mathsf{TYPE}(v_2)=\RUD 
                     \end{array} }\\\\
%
\end{array}\]
}
\justifying

\noindent
These rules mean that, for any $\mathit{Bop}=\{\wedge, \vee \}$,
if one operand is $\RUD$, the other operand is $\SID$, and they share
no input, then $v$ has a secret independent distribution ($\SID$).
\emph{GMul} denotes  multiplication 
in a finite field.
Here, $\mathsf{supp}(v_1) \cap\mathsf{supp}(v_2)=\emptyset$ is need;
otherwise, the common input may cause problem. 
For example, if $v_1 \leftarrow m \oplus k$ and $v_2 \leftarrow
m \wedge x$, then $v = (v_1\wedge v_2) = (m\wedge\neg k) \wedge x$ has
a leak because if $k=1$, $v=0$; but if $k=0$, $v=m\wedge x$.

{\footnotesize
\[
  {Rule}_{4} \myinferrule
                   {\mathsf{TYPE}(v) = \SID}
                   {\begin{array}{c}
                    v\leftarrow \mathit{Bop}(v_1,v_2) \wedge 
                    \mathsf{supp}(v_1)\cap\mathsf{supp}(v_2) = \emptyset \\   
                    \wedge
                    \mathsf{TYPE}(v_1)=\SID \wedge
                    \mathsf{TYPE}(v_2)=\SID
                    \end{array}}
\]
}

\noindent
Similarly, $\mathit{Rule_4}$ may elevate a variable $v$ from $\UKD$ to
$\SID$, e.g., as in $v\leftarrow ((k\oplus m) \wedge x_1) \wedge
(x_2)$ where $x_1$ and $x_2$ are both $\SID$.  
Again, the condition
$\mathsf{supp}(v_1)\cap\mathsf{supp}(v_2) = \emptyset$ in
$\mathit{Rule_4}$ is needed because, otherwise, there may be cases
such as $v \leftarrow ((k\oplus m)\wedge x_1) \wedge (x_2\wedge m)$,
which is equivalent to $v \leftarrow \neg k\wedge (m \wedge x_1 \wedge
x_2)$ and thus has a leak.

Figure~\ref{inference} shows the other inference rules used in our
system.  Since these rules are self-explanatory, we omit the proofs.
%

\begin{figure*}
\noindent\begin{minipage}{.45\textwidth}
{\footnotesize
\[
  {Rule}_{5a} \myinferrule
                   {\mathsf{TYPE}(v) = \SID}
                   { \begin{array}{c}
                    v\leftarrow \mathit{Bop}(v_1,v_2)    
                     \wedge						\\
                    \mathsf{dom}(v_1)\setminus\mathsf{supp}(v_2) = \emptyset
                    \wedge
                     \mathsf{TYPE}(v_1)=\RUD 
                     \wedge 						\\
                     \mathsf{dom}(v_1) =  \mathsf{dom}(v_2) 
                    \wedge 
                     \mathsf{supp}(v_1) =  \mathsf{supp}(v_2) 						
                    \end{array} }
\]
}
\end{minipage}
\begin{minipage}{.45\textwidth}
{\footnotesize
\[
  {Rule}_{5b} \myinferrule
                   {\mathsf{TYPE}(v) = \SID}
                   { \begin{array}{c}
                    v\leftarrow \mathit{Bop}(v_1,v_2)    
                     \wedge						\\
                    \mathsf{dom}(v_2)\setminus\mathsf{supp}(v_1) = \emptyset
                    \wedge
                     \mathsf{TYPE}(v_2)=\RUD 
                     \wedge 						\\
                     \mathsf{dom}(v_1) =  \mathsf{dom}(v_2) 
                    \wedge 
                     \mathsf{supp}(v_1) =  \mathsf{supp}(v_2) 						
                    \end{array} }
\]
}
\end{minipage}
\vspace{-.5em}
\begin{minipage}{.95\textwidth}
{\footnotesize
\[
  {Rule}_{6} \myinferrule
                   {\mathsf{TYPE}(v) = \SID}
                   { \begin{array}{c}
                    v\leftarrow \mathit{Bop}(v_1,v_2) \wedge 
                    \mathit{Bop} \not\in \{\mathit{Xor}, \mathit{GMul} \} \wedge 
                  (\mathsf{dom}(v_1)\setminus \mathsf{supp}(v_2) \neq \emptyset \vee
                  \mathsf{dom}(v_2)\setminus \mathsf{supp}(v_1) \neq \emptyset )
                    \wedge
                    \mathsf{TYPE}(v_1)=\RUD \wedge
                    \mathsf{TYPE}(v_2)=\RUD
                    \end{array} }
\]
}
\end{minipage}
\vspace{-.5em}
\begin{minipage}{.45\textwidth}
{\footnotesize
\[
  {Rule}_{7a} \myinferrule
                   {\mathsf{TYPE}(v) = \SID}
                   { \begin{array}{c}
                    v\leftarrow \mathit{Bop}(v_1,v_2) \wedge 
                    \mathit{Bop} = \mathit{GMul}  \wedge 
                    \mathsf{TYPE}(v_1) = \RUD \wedge
                    \\
                    \mathsf{TYPE}(v_2) = \SID \wedge
                     \mathsf{dom}(v_1) \setminus \mathit{supp}(v_2) \neq \emptyset
                    \end{array} }
\]
}
\end{minipage}
\begin{minipage}{.45\textwidth}
{\footnotesize
\[
  {Rule}_{7b} \myinferrule
                   {\mathsf{TYPE}(v) = \SID}
                   { \begin{array}{c}
                    v\leftarrow \mathit{Bop}(v_1,v_2) \wedge 
                    \mathit{Bop} = \mathit{GMul}  \wedge 
                    \mathsf{TYPE}(v_1) = \SID \wedge
                    \\
                    \mathsf{TYPE}(v_2) = \RUD \wedge 
                    \mathsf{dom}(v_2) \setminus \mathit{supp}(v_1) \neq \emptyset
                    \end{array} }
\]
}
\end{minipage}
\vspace{-.5em}
\begin{minipage}{.95\textwidth}
{\footnotesize
\[
  {Rule}_{8} \myinferrule
                   {\mathsf{TYPE}(v) = \SID}
                   { \begin{array}{c}
                    v\leftarrow \mathit{Bop}(v_1,v_2) \wedge 
                    \mathit{Bop} = \mathit{GMul}  \wedge 
                    (\mathsf{dom}(v_1) \setminus \mathit{dom}(v_2) \neq \emptyset \vee
                    \mathsf{dom}(v_2) \setminus \mathit{dom}(v_1) \neq \emptyset)  \wedge
                    \mathsf{TYPE}(v_1) = \RUD \wedge
                    \mathsf{TYPE}(v_2) = \RUD 
                    \end{array} }
\]
}
\end{minipage}



\vspace{-2ex}
\caption{The remaining inference rules used in our type system (in addition to $\mathit{Rule}_{1-4}$).}
\label{inference}
\end{figure*}

\subsection{Detecting HD-sensitive Pairs}
\label{sec:detecting-HD-sensitive-pairs}

Based on the variable types, we compute HD-sensitive pairs. For each
pair $(v_1,v_2)$, we check if $HD(v_1,v_2)$ results in a leak when
$v_1$ and $v_2$ share a register.  There are two scenarios:
\begin{itemize}
\item
$v_1 \leftarrow \mathit{expr_1}; v_2 \leftarrow \mathit{expr_2}$,
meaning $v_1$ and $v_2$ are defined in two instructions.
\item
$v_1 \leftarrow \mathit{Bop} (v_2, v_3)$, where the result $v_1$ and
one operand $v_2$ are stored in the same register.
\end{itemize}
In the \emph{two-instruction} case, we check $HW(expr_1\oplus expr_2)$
using \emph{Xor}-related inference rules. 
For example, if $v_1 \leftarrow k\oplus m$ and $v_2 \leftarrow m$,
since $m$ appears in the supports of both expressions, $(k\oplus
m)\oplus m$ is $\UKD$.
Such leak will be denoted $\hdsenDouble{}(v_1,v_2)$, where $D$ stands
for ``Double''.

In the \emph{single-instruction} case, we check
$HW( \mathit{Bop}(v_2,v_3) \oplus v_2)$ based on the operator type.
When $\mathit{Bop} = \wedge$, we have $(v_2 \wedge v_3)\oplus v_2 =
v_2 \wedge \neg v_3$;
when $\mathit{Bop} = \vee$, we have $(v_2 \vee v_3) \oplus v_2 = (\neg
v_2 \wedge v_3)$;
when $\mathit{Bop} = \oplus$ (\emph{Xor}), we have $(v_2 \oplus
v_3) \oplus v_2 = v_3$; and
when $\mathit{Bop} = \ast$ (\emph{GMul}), the result of $(v_2 \ast
v_3) \oplus v_2$ is $\{v_2, v_3\}$ if $v_2 \ast v_3 \neq 0x01$ and is
$(v_2 \oplus 0x01)$ otherwise.  
Since the type inference procedure is agnostic to the result of
$(v_2 \ast v_3)$, the type of $(v_2 \ast v_3) \oplus v_2$ depends on
the types of $v_3$ and $v_2$; that is, $ \mathsf{TYPE}(v_2)
= \UKD \vee
\mathsf{TYPE}(v_3) = \UKD \implies \mathsf{TYPE}((v_2 \ast v_3)
 \oplus v_2) = \UKD$.
%
%
%
If there is a leak, it will be denoted $\hdsenSingle{}(v_1,v_2)$.

The reason why HD leaks are divided to $\hdsenDouble{}$ and
$\hdsenSingle{}$ is because they have to be mitigated differently.
When the leak involves two instructions, it may be mitigated by
constraining the register allocation algorithm such that $v_1$ and
$v_2$ no longer can share a register. 
In contrast, when the leak involves a single instruction, it cannot be
mitigated in this manner because in x86, for example, all binary
instructions require the result to share the same register or memory
location with one of the operands.  Thus, mitigating the $\hdsenSingle{}$ 
requires that we rewrite the instruction itself.

\vspace{2mm}
We also define a relation $\mathit{Share}(v_1,v_2)$, meaning $v_1$ and
$v_2$ indeed may share a register, and use it to filter the
HD-sensitive pairs, as shown in the two rules below.  
\vspace{2ex}
%
{\footnotesize
\[\begin{array}{l}
        \myinferrule {\hdsenDouble{}(v_1,v_2)}
                    {\mathit{Share}(v_1,v_2) \wedge \mathsf{TYPE}(v_1\oplus v_2) = \UKD 
                     \wedge v_1 \leftarrow expr_1 \wedge v_2 \leftarrow expr_2
                    }
\end{array}\]
\[\begin{array}{l}
        \myinferrule{\hdsenSingle{}(v_1,v_2)}
                    {\mathit{Share}(v_1,v_2) \wedge \mathsf{TYPE}(v_1\oplus v_2) = \UKD 
                     \wedge v_1 \leftarrow Bop(v_2,v_3)
                    }
\end{array}\]

}
\noindent

Backend information (Section~\ref{sec:backendinfo}) is required to
define the relation; for now, assume $\forall v_1,v_2:
\mathit{Share}(v_1,v_2) = \mathit{true}$.

\vspace{.5mm}
\section{Mitigation during Code Generation}
\label{sec:mitigation}

We mitigate leaks by using the two types of HD-sensitive pairs as
constraints during register allocation.

\paragraph{Register Allocation}

The classic approach, especially for static compilation, is based on
\emph{graph coloring}~\cite{chaitin2004register,GeorgeA96}, whereas
dynamic compilation may use faster algorithms such as \emph{lossy graph
coloring}~\cite{CooperD06} or \emph{linear
scan}~\cite{PolettoS99}.
We apply mitigation on both graph coloring and LLVM's basic register
allocation algorithms. For ease of comprehension, we use graph
coloring to illustrate our constraints.

In graph coloring, each variable corresponds to a node and each edge
corresponds to an \emph{interference} between two variables, i.e.,
they may be in use at the same time and thus cannot occupy the same
register.  Assigning variables to $k$ registers is similar to coloring
the graph with $k$ colors.
To be efficient, variables may be grouped to clusters,
or \emph{virtual registers}, before they are assigned to physical
registers (colors).  In this case, each virtual register
(\emph{vreg}), as opposed to each variable, corresponds to a node in
the graph, and multiple virtual registers may be mapped to one physical
register.

\subsection{Handling $\hdsenDouble{}$ Pairs}

For each $\hdsenDouble{}(v_1,v_2)$, where $v_1$ and $v_2$ are defined
in two instructions, we add the following constraints. 
First, $v_1$ and $v_2$ are not to be mapped to the same virtual
register.
Second, virtual registers $\mathit{vreg_1}$ and $\mathit{vreg_2}$ (for
$v_1$ and $v_2$) are not to be mapped to the same physical register.
Toward this end, we constrain the behavior of two backend
modules: 
 \emph{Register Coalescer} and \emph{Register Allocator}.

%
Our constraint on \emph{Register Coalescer} states that $vreg_1$ and
$vreg_2$, which correspond to $v_1$ and $v_2$, must never coalesce,
although each of them may still coalesce with other virtual registers.
As for \emph{Register Allocator}, our constraint is on the formulation
of the graph. For each HD-sensitive pair, we add a
new \emph{interference} edge to indicate that $vreg_1$ and $vreg_2$
must be assigned different colors.

During graph coloring, these new edges are treated the same as all
other edges.  Therefore, our constraints are added to the register
allocator and its impact is propagated automatically to all subsequent
modules, regardless of the architecture (x86, MIPS or ARM).
When variables cannot fit in the registers, some will
be \emph{spilled} to memory, and all reference to them will be
directed to memory.
Due to the constraints we added, there may be more spilled variables,
but spilling is handled transparently by the existing algorithms in
LLVM. 
This is an advantage of our approach: it identifies a way to constrain
the behavior of existing modules in LLVM, without the need to
rewriting any of them from scratch.

\subsection{Handling $\hdsenSingle{}$ Pairs}

For each $\hdsenSingle{}(v_1,v_2)$ pair, where $v_1$ and $v_2$ appear
in the same instruction, we additionally constrain the \emph{DAG
Combiner} module to rewrite the instruction before constraining the
register allocation modules.  To see why, consider $mk = (m \oplus
k)$, which compiles to

\vspace{1ex}
\hspace{4ex}
\inlinecode{C}{MOVL -4(\%rbp), \%ecx // -4(\%rbp) =  m (random)}

\vspace{-0.5ex}
\hspace{4ex}
\inlinecode{C}{XORL -8(\%rbp), \%ecx // -8(\%rbp) =  k (secret)}
\vspace{1ex}

\noindent
Here, -4(\%rbp) and -8(\%rbp) are memory locations for $m$ and $k$,
respectively.  Although $m$ and $mk$ are  $\RUD$ (no leak) when
stored in \%ecx, the transition from $m$ to $mk$, $HW(m\oplus mk) =
k$, has a leak.

To remove the leak, we must rewrite the instruction:

\vspace{1ex}
\hspace{4ex}
\inlinecode{C}{MOVL -4(\%rbp), \%ecx   // -4(\%rbp)= m}

\vspace{-0.5ex}
\hspace{4ex}
\inlinecode{C}{XORL \%ecx, -8(\%rbp)   		// -8(\%rbp)= k, and then mk}
\vspace{1ex}

\noindent
While $m$ still resides in \%ecx, both $k$ and $mk$ reside in the
memory -8(\%rbp).
There is no leak because \%ecx only stores $m$ ($\RUD$) and
$HW(m \oplus m) = 0$.
Furthermore, the solution is efficient in that no additional memory is
needed.
If $k$ were to be used subsequently, we would copy $k$ to another
memory location and re-directed uses of $k$ to that location.

\paragraph{Example}

Figure~\ref{maskExample} shows a real program~\cite{yao2018fault},
where $s$ is an array storing sensitive data while $m1$-$m8$ are
random masks.  The compiled code (left) has leaks,
whereas the mitigated code (right) is leak free.
The reason why the original code (left) has leaks is because, prior
to Line~8, \%eax stores $m1 \oplus m5$, whereas after Line~8, 
\%eax stores $s[0+i*4] \oplus m1 \oplus m5$; thus, bit-flips in 
\%eax is reflected in $HW({\%eax}_{1}\oplus{\%eax}_{2}) = s[0+i*4]$, 
which is the sensitive data.

\begin{figure}

\centering
\begin{minipage}{.46\textwidth}
\begin{lstlisting}[numbers=left,numberstyle=\tiny, basicstyle=\ttfamily\scriptsize]
void remask(uint8_t s[16], uint8_t m1, uint8_t m2, uint8_t m3, uint8_t m4, uint8_t m5, uint8_t m6, uint8_t m7, uint8_t m8){
	int i;	
	for(i = 0; i< 4; i++){
		s[0+i*4] = s[0+i*4] ^ (m1^m5);
		s[1+i*4] = s[1+i*4] ^ (m2^m6);
		s[2+i*4] = s[2+i*4] ^ (m3^m7);
		s[3+i*4] = s[3+i*4] ^ (m4^m8);
	}
}
\end{lstlisting}
\end{minipage}

\begin{minipage}{.21\textwidth}
\begin{lstlisting}[firstnumber=1, frame=tlrb, numbers=left,numberstyle=\tiny, basicstyle=\ttfamily\scriptsize]{Name}
	//Before Mitigation							
	movslq	-28(%rbp), %rdx							
	movq	-16(%rbp), %rcx							
	movzbl	(%rcx,%rdx,4), %edi			
	movzbl	-17(%rbp), %esi						
	movzbl	-21(%rbp), %eax						
	xorl	%esi, %eax									
	xorl	%edi, %eax						
	movb	%al, (%rcx,%rdx,4)
\end{lstlisting}
\end{minipage}
\hspace{1.5em}
\begin{minipage}{.21\textwidth}
\begin{lstlisting}[firstnumber=1, frame=tlrb, numbers=left,numberstyle=\tiny, basicstyle=\ttfamily\scriptsize]{Name}
	//After mitigation
	movslq	-28(%rbp), %rdx
	movq	-16(%rbp), %rcx

	movzbl	-17(%rbp), %esi
	movzbl	-21(%rbp), %eax
	xorl	%esi, %eax

	xorb	%al, (%rcx,%rdx,4)
\end{lstlisting}
\end{minipage}
\vspace{-2ex}
\caption{Code snippet from the Byte Masked AES~\cite{yao2018fault}.}

\label{maskExample}
\end{figure}

During register allocation, a virtual register $\mathit{vreg_1}$ would
correspond to $m1 \oplus m5$ while $\mathit{vreg_2}$ would correspond
to $s[0+i*4] \oplus m1 \oplus m5$.  Due to a constraint from
this \hdsenSingle{} pair, our method would prevent $\mathit{vreg_1}$
and $\mathit{vreg_2}$ from coalescing, or sharing a physical
register. After rewriting, $\mathit{vreg_2}$ shares the same memory
location as $s[0+i*4]$ while $\mathit{vreg_1}$ remains unchanged.
Thus, $m1 \oplus m5$ is stored in \%al and $s[0+i*4] \oplus m1 \oplus
m5$ is spilled to memory, which removes the leak.

\section{Domain-specific Optimizations}
\label{sec:optimization}

While the method presented so far has all the functionality, it can be
made faster by domain-specific optimizations.

\subsection{Leveraging the Backend Information}
\label{sec:backendinfo}

To detect HD leaks that likely occur, we focus on pairs of variables
that may share a register as opposed to arbitrary pairs of variables.
For example, if the live ranges of two variables overlap, they will
never share a register, and we should not check them for HD leaks.
Such information is readily available in the compiler's backend
modules, e.g., in graph coloring based register allocation, variables
associated with any \emph{interference} edge cannot share a register.

Thus, we define $\mathit{Share}(v_1,v_2)$, meaning $v_1$ and $v_2$ may
share a register.
After inferring the variable types as $\RUD$, $\SID$, or $\UKD$, we
use $\mathit{Share}(v_1,v_2)$ to filter the variable pairs subjected
to checking for \hdsenDouble{} and \hdsenSingle{} leaks (see
Section~\ref{sec:detecting-HD-sensitive-pairs}).
%
%
We will show in experiments that such backend information allows us to
dramatically shrink the number of HD-sensitive pairs.

\subsection{Pre-computing Datalog Facts}
\label{sec:precomputation}

By default, only input annotation and basic data-flow (def-use) are
encoded as Datalog facts, whereas the rest has to be deduced by
inference rules.
However, Datalog is not the most efficient way of computing sets, such
as $\mathit{supp}(v)$, $\mathit{unq}(v)$ and $\mathit{dom}(v)$, or
performing set operations such as $m_1\in\mathit{supp}(v)$.

In contrast, it is linear time~\cite{OMHE17} to compute sets such as
$\mathit{supp}(v)$, $\mathit{unq}(v)$ and $\mathit{dom}(v)$
explicitly.
Thus, we choose to precompute them in advance and encode the results
as Datalog facts.  In this case, precomputation results are used to
jump start Datalog based type inference.  We will show, through
experiments, that the optimization can lead to faster type inference
than the default implementation.

\subsection{Efficient Encoding of Datalog Relations}
\label{sec:encoding}

There are different encoding schemes for Datalog.  For example, if
$\mathit{IN} =\{i_0,\ldots,i_3\}$ and $\mathit{supp(v_1)}=\{i_1,i_2\}$
and $\mathit{supp(v_2)} = \{i_0,i_1,i_3\}$.
One way is to encode the sets is using a relation $\mathit{Supp:
V\times IN}$, where $V$ are variables and $\mathit{IN}$ are inputs:
\[\begin{array}{ll}
  \mathit{Supp(v_1,i_1) \wedge Supp(v_1,i_2)}                      & = \mathit{supp(v_1)} \\
  \mathit{Supp(v_2,i_0) \wedge Supp(v_2,i_1) \wedge Supp(v_2,i_3)}  & = \mathit{supp(v_2)} \\
\end{array}\]
While the size of $\mathit{Supp}$ is $|V||IN|$, each set needs up to
$|IN|$ predicates, and set operation needs $|IN|^2$ predicates.

Another way is to encode the sets is using a relation $\mathit{Supp}:
V\times 2^{IN}$, where $2^\mathit{IN}$ is the power-set (set of all
subsets of $\mathit{IN}$):
\[\begin{array}{ll}
  \mathit{Supp}(v_1,b0110)    & = \mathit{supp(v_1)}  \\
  \mathit{Supp}(v_2,b1011)    & = \mathit{supp(v_2)}  \\
\end{array}\]
While the size of \emph{Supp} is $|V|~ 2^{|IN|}$, each set needs 
one predicate, and set operation needs 2 predicates (a bit-wise
operation).  When $|IN|$ is small, the second approach is
more compact; but as $|IN|$ increases, the table size
of \emph{Supp} increases exponentially.

Therefore, we propose an encoding, called segmented bitset
representation \emph{(idx,bitset)}, where \emph{idx=i} refers to the
$i$-th segment and \emph{bitset$_i$} denotes the bits in the $i$-th
segment.
\[\begin{array}{ll}
  \mathit{Supp(v_1,1, b01) \wedge Supp(v_1, 0, b10)} & = \mathit{supp(v_1)}   \\
  \mathit{Supp(v_2,1, b10) \wedge Supp(v_2, 0, b11)} & = \mathit{supp(v_2)}   \\
\end{array}\]
In practice, when the \emph{bitset} size is bounded, e.g., to 4, the table size remains
small while the number of predicates increases moderately.  
This encoding scheme is actually a generalization of the previous two.
When the size of \emph{bitset} decreases  
to 1 and the number of segments increases
to $|IN|$, it degenerates to the first approach.  When the size of \emph{bitset}
increases to $|IN|$ and the number of segments decrease to 1, it
degenerates to the second approach.

\begin{table}
\caption{Statistics of the benchmark programs.}
\vspace{-2ex}
\label{tbl:stats}
\centering
\scalebox{0.61}{
\begin{tabular}{|l|l|r|c|c|c|r|}
\hline
\multirow{2}{*}{Name} & \multirow{2}{*}{Description} & \multirow{2}{*}{LoC} & \multicolumn{4}{c|}{Program  Variables}  \\ \cline{4-7}
    &                                               &           & $\mathit{IN_{PUBLIC}}$ & $\mathit{IN_{SECRET}}$ & $\mathit{IN_{RANDOM}}$   &  Internal     \\\hline\hline

P1  & AES Shift Rows~\cite{bayrak2013sleuth}                   & 11        &0          &2          &2          &22                \\ \hline
P2  & Messerges Boolean~\cite{bayrak2013sleuth}                & 12        &0          &2          &2          &23                \\ \hline
P3  & Goubin Boolean~\cite{bayrak2013sleuth}                   & 12        &0          &1          &2          &32                \\ \hline
P4  & SecMultOpt\_wires\_1~\cite{rivain2010provably}             & 25        &1          &1          &3          &44                \\ \hline
P5  & SecMult\_wires\_1~\cite{rivain2010provably}                & 25        &1          &1          &3          &35                \\ \hline
P6  & SecMultLinear\_wires\_1~\cite{rivain2010provably}          & 32        &1          &1          &3          &59                \\ \hline
P7  & CPRR13-lut\_wires\_1~\cite{coron2013higher}             & 81        &1          &1          &7          &169                \\ \hline
P8  & CPRR13-OptLUT\_wires\_1~\cite{coron2013higher}          & 84        &1          &1          &7          & 286               \\ \hline
P9  & CPRR13-1\_wires\_1~\cite{coron2013higher}               & 104       &1          &1          &7          &207                \\ \hline
P10 & KS\_transitions\_1~\cite{barthe2015verified}               & 964       &1          &16          &32          &2,329                \\ \hline
P11 & KS\_wires~\cite{barthe2015verified}                        & 1,130     &1          &16          &32          &2,316               \\ \hline
P12 & keccakf\_1turn~\cite{barthe2015verified}                   & 1,256     &0          &25          &75          &2,314                \\ \hline
P13 & keccakf\_2turn~\cite{barthe2015verified}                   & 2,506     &0          &25          &125          &4,529                \\ \hline
P14 & keccakf\_3turn~\cite{barthe2015verified}                   & 3,764     &0          &25          &175          &6,744                \\ \hline
P15 & keccakf\_7turn~\cite{barthe2015verified}                   & 8,810     &0          &25          &349          &15,636                \\ \hline
P16 & keccakf\_11turn~\cite{barthe2015verified}                  & 13,810    &0          &25          &575          &24,472                \\ \hline
P17 & keccakf\_15turn~\cite{barthe2015verified}                  & 18,858    &0          &25          &775          &33,336                \\ \hline
P18 & keccakf\_19turn~\cite{barthe2015verified}                  & 23,912    &0          &25          &975          &42,196                \\ \hline
P19 & keccakf\_24turn~\cite{barthe2015verified}                  & 30,228    &0          &25          &1,225          &53,279                \\ \hline
P20 & AES\_wires\_1~\cite{coron2013higher}                    & 34,358    &16          &16          &1,232          &63,263                \\ \hline

\end{tabular}
}
\end{table}

\section{Experiments}
\label{sec:experiments}

We have implemented our method in LLVM 3.6~\cite{lattner2004llvm}. It
leverages the $\mu${Z}~\cite{Hoder11} Datalog engine in
Z3~\cite{DeMoura08} to infer types, and then constrains backend
modules using the inferred HD leaks.  While the mitigation part is
implemented specifically for x86 in LLVM, this approach can also be
implemented for other platforms.

We conducted experiments on a number of cryptographic software
benchmarks.
%
%
Table~\ref{tbl:stats} shows the statistics, including the name, a
short description, the number of lines of code (LoC), and the number of
program variables, which are divided further to input and internal
variables.
All benchmarks are from public domain, and all of them are masked.
The programs P1-P3, in particular, are protected by Boolean masking
that previously has been
verified~\cite{bayrak2013sleuth,eldib2014formal,zhang2018scinfer}.
The other programs, from Barthe et al.~\cite{barthe2015verified}, are
masked multiplication~\cite{rivain2010provably}, masked
S-box~\cite{coron2013higher}, masked AES~\cite{coron2013higher} and various 
masked MAC-Keccak functions \cite{barthe2015verified}.

Our experiments were designed to answer three questions:
(1) Is our Datalog-based type system effective in detecting HD leaks?
(2) Are the domain-specific optimizations effective in reducing the
computational overhead?
(3) Does the mitigated code have good performance after compilation,
in terms of both the code size and the execution speed?

In all the experiments, we used a computer with 2.9 GHz CPU and 8GB
RAM, and set the timeout (T/O) to 120 minutes.

\subsection{Leak Detection Results}

Table~\ref{tbl:detection} shows the results, where Columns 1-2 show
the benchmark name and detection time and Columns 3-4 show the number
of HD leaks detected.  The leaks are further divided
into \hdsenDouble{} (two-instruction) and \hdsenSingle{}
(single-instruction).  Columns~5-7 show more details of the type
inference, including the number of $\RUD$, $\SID$ and $\UKD$
variables, respectively.
While the time taken to complete type inference is not negligible,
e.g., minutes for the larger programs, it is reasonable because we
perform a much deeper program analysis than mere compilation.  To put
it into perspective, the heavy-weight formal verification
approaches often take hours~\cite{eldib2014formal,zhang2018scinfer}.

As for the number of leaks detected, although the benchmark programs
are all masked, during normal compilation, new HD leaks were still
introduced as a result of register reuse.  For example, in P20, which
is a masked AES~\cite{barthe2015verified}, we detected 33 \hdsenSingle{} leaks
after analyzing more than 60K intermediate variables.  Overall, we
detected HD leaks in 17 out of the 20 programs. Furthermore, 6 of
these 17 programs have both \hdsenDouble{} and \hdsenSingle{} leaks,
while the remaining 11 have only \hdsenSingle{} leaks.

Results in columns 5-7 of Table~\ref{tbl:detection}
indicate the inferred types of Program Variables.
Despite the large number of 
variables in a program, our type inference method did an excellent job
in quickly proving that the vast majority of them are $\RUD$ or
$\SID$ (no leak); even for the few $\UKD$ variables, after the
backend information is used, the number of actual HD leaks detected by
our method is  small.

\begin{table}
\caption{Results of type-based HD leak detection.}
\vspace{-2ex}
\label{tbl:detection}
\centering
\scalebox{0.61}{
\begin{tabular}{|l||r||c|c||c|c|c|}
\hline
\multirow{2}{*}{Name} & \multirow{2}{*}{Detection Time} & \multicolumn{2}{c||}{HD Leaks Detected} & \multicolumn{3}{c|}{Details of the Inferred Types} \\\cline{3-7}
     &                &\ \ \ \ $\hdsenDouble{}$\ \ \  &\ \ \ \ $\hdsenSingle{}$\ \ \  & $\RUD$   & $\SID$   & $\UKD$    \\\hline\hline

P1   & 0.061s         &  NONE          & NONE            & 22     & 0        & 4        \\\hline
P2   & 0.105s         &  NONE          & NONE            & 20     & 0        & 7        \\\hline
P3   & 0.099s         &  NONE          & 2               & 31     & 3        & 1        \\\hline
P4   & 0.208s         &  NONE          & 2               & 31     & 12       & 6        \\\hline
P5   & 0.216s         &  NONE          & 2               & 29     & 10       & 1        \\\hline
P6   & 0.276s         &  4             & 2               & 48     & 15       & 1        \\\hline
P7   & 0.213s         &  10            & 2               & 151    & 25       & 2        \\\hline
P8   & 0.147s         &  12            & 2               & 249    & 42       & 4        \\\hline
P9   & 0.266s         &  6             & 2               & 153    & 61       & 2        \\\hline
P10  & 0.550s         &  NONE             & NONE              & 2,334  & 12       & 31       \\\hline
P11  & 0.447s         &  4          & 16            & 2,334  & 0        & 31       \\\hline
P12  & 0.619s         &  NONE          & 7               & 2,062  & 300      & 52       \\\hline
P13  & 1.102s         &  NONE          & 5               & 4,030  & 600      & 49       \\\hline
P14  & 1.998s         &  NONE          & 5               & 5,995  & 900      & 49       \\\hline
P15  & 16.999s        &  NONE          & 25              & 13,861 & 2,100    & 49       \\\hline
P16  & 24.801s        &  NONE          & 5               & 21,723 & 3,300    & 49       \\\hline
P17  & 59.120s        &  NONE          & 5               & 29,587 & 4,500    & 49       \\\hline
P18  & 2m1.540s       &  NONE          & 4               & 37,449 & 5,700    & 47       \\\hline
P19  & 3m22.415s      &  NONE          & 5               & 47,280 & 7,200    & 49       \\\hline
P20  & 16m12.320s     &  29            & 33              & 38,070 & 26,330 & 127      \\\hline

\end{tabular}
}
\end{table}

\subsection{Effectiveness of Optimizations}

To quantify the impact of our optimizations, we measured the
performance of our method with and without them.
Table~\ref{tbl:optimization} shows the significant differences in
analysis time (Columns 2-3) and detected HD leaks (Columns 4-7).  
Overall, the optimized version completed all benchmarks whereas the
unoptimized only completed half.  
For P12, in particular, the
optimized version (Section 6) was 11,631X faster.
In unoptimized version, since the memory requirement increases
as the |$IN$| goes up, P12 ran out of memory and started using 
virtual memory, which resulted in the slow-down.



Leveraging the backend information also drastically reduced the number
of detected leaks.
This is because, otherwise, we have to be conservative and
assume any two variables may share a register, which causes 
a lot of false leaks in X86 platform.
In P12, for example,
using the backend information resulted in 260X fewer leaks.

\begin{table}
\caption{Results of quantifying impact of optimizations.}
\vspace{-2ex}
\label{tbl:optimization}
\centering
\scalebox{0.61}{
\begin{tabular}{|l||r|r||c|c|c|c|}
\hline
\multirow{2}{*}{Name} & \multicolumn{2}{c||}{Detection Time} & \multicolumn{2}{c|}{Without Backend-Info} &\multicolumn{2}{c|}{With Backend-Info}  \\\cline{2-7}
     &  w/o optimization     &  w/ optimization  & \hdsenDouble{} & \hdsenSingle{} & \hdsenDouble{} & \hdsenSingle{}    \\ \hline\hline

P1   &0.865s        & 0.061s   & 0      & 18      & 0     & 0      \\ \hline
P2   &0.782s        & 0.105s   & 0      & 9       & 0     & 0      \\ \hline
P3   &0.721s        & 0.099s   & 0      & 15      & 0     & 2      \\ \hline
P4   &1.102s        & 0.208s   & 0      & 32      &  0    & 2      \\ \hline
P5   &1.206s        & 0.216s   & 0      & 32      &  0    & 2      \\ \hline
P6   &1.113s        & 0.276s   & 8      & 40      &  4    & 2      \\ \hline
P7   &5.832s        & 0.213s   & 44     & 144     &  10    & 2      \\ \hline
P8   &4.306s        & 0.147s   & 68     & 323     &  12    & 2      \\ \hline
P9   &5.053s        & 0.266s   & 43     & 160     &  6    & 2      \\ \hline
P10  &10m1.513s          & 0.550s   & 12     & 180     &  0    & 0     \\ \hline
P11  &15m51.969s          & 0.447s   & 12     & 180     &  4    & 16     \\ \hline
P12  &T/O        & 0.619s   & 473    & 1,820   &  0    & 7      \\ \hline
P13  &T/O           & 1.102s   & 492    & 1,884   &  0    & 5      \\ \hline
P14  &T/O           & 1.998s   & 492    & 1,884   &  0    & 5      \\ \hline
P15  &T/O           & 16.999s  & 492    & 1,884   &  0    & 25     \\ \hline
P16  &T/O           & 24.801s  & 492    & 1,884   &  0    & 5      \\ \hline
P17  &T/O           & 59.120s  & 492    & 1,884   &  0    & 5      \\ \hline
P18  &T/O           & 2m1s     & 468    & 1,800   &  0    & 4      \\ \hline
P19  &T/O           & 3m22s    & 492    & 1,884   &  0    & 5      \\ \hline
P20  &T/O           & 16m13s   & 620    & 1,944   & 29    & 33     \\ \hline

\end{tabular}
}
\end{table}

\subsection{Leak Mitigation Results}

We compared the size and execution speed of the LLVM compiled code,
with and without our mitigation.
The results are shown in Table~\ref{tbl:mitigation}, including the
number of bytes in the assembly code and the execution
time. 
Columns~8-9 show more details: the number of virtual registers
marked as sensitive and non-sensitive, respectively.

The results show that our mitigation has little performance overhead.
First, the code sizes are almost the same.  For P8, the mitigated code
is even smaller because, while switching the storage from register to
memory during our handling of the \hdsenSingle{} pairs, subsequent
memory stores may be avoided.
Second, the execution speeds are also similar. 
Overall, the mitigated
code is 8\%-11\% slower, but in some cases, e.g., P4 and P6, the
mitigated code is faster because of our memory related rewriting.

The main reason why our mitigation has little performance overhead is
because, as shown in the last two columns of
Table~\ref{tbl:mitigation}, compared to the total number of virtual
registers, the number of sensitive ones is extremely small.
P17 (keccakf\_15turn), for example, has only 5 sensitive virtual
registers out of the 9,917 in total.
Thus, our mitigation only has to modify a small percentage of the
instructions, which does not lead to significant overhead.

\begin{table}
\caption{Results of our HD leak mitigation.}
\vspace{-2ex}
\label{tbl:mitigation}        
\centering
\scalebox{0.61}{
\begin{tabular}{|l||r|r|c||r|r|c||r|r|}
\hline
\multirow{2}{*}{Name} & \multicolumn{3}{c||}{Code-size Overhead (byte)} & \multicolumn{3}{c||}{Runtime Overhead (us)} & \multicolumn{2}{c|}{Virtual Register} \\ \cline{2-9} 
    & original & mitigated& \%     & original&mitigated& \%       & sensitive    & non-sensitive \\ \hline

P3  & 858      & 855      &0.3        &  -      &  -      &  -        & 2           & 4             \\ \hline
P4  & 1,198    & 1,174    &2        & 0.23    & 0.20    & -13         & 2           & 13            \\ \hline
P5  & 1,132    & 1,108    &2.12        & 0.30    & 0.37    & 2.3         & 2           & 9             \\ \hline
P6  & 1,346    & 1,339    &0.52        & 0.30    & 0.27    & -10         & 5           & 8             \\ \hline
P7  & 3,277    & 3,223    &1.64        & 0.29    & 0.30    &  3.4        & 10          & 27            \\ \hline
P8  & 3,295    & 3,267    &0.85        & 0.20    & 0.22    & 10         & 11          & 83            \\ \hline
P9  & 3,725    & 3,699    &0.69        & 0.7    & 0.78    & 11         & 10          & 29            \\ \hline
P11 & 44,829   & 44,735   &0.21        & 5.60    & 6.00    & 7.1         & 18          & 680           \\ \hline
P12 & 46,805   & 46,787   &0.03        & 6.20    & 6.50    &  4.83        & 7           & 726           \\ \hline
P13 & 90,417   & 90,288   &0.14        & 13.60   & 13.00   & -4.41         & 5           & 1,384         \\ \hline
P14 & 134,060  & 133,931  &0.09        & 23.00   & 21.00   & -8.69         & 5           & 2,040         \\ \hline
P15 & 313,454  & 312,930  &0.16        & 52.00   & 58.00   & 11.5         & 25          & 4,637         \\ \hline
P16 & 496,087  & 495,943  &0.03        & 91.00   & 96.00   & 5.49         & 5           & 7,288         \\ \hline
P17 & 677,594  & 677,450  &0.02        & 129.00  & 136.00  &  5.42        & 5           & 9,912         \\ \hline
P18 & 859,150  & 859,070  &0.009        & 178.00  & 183.00  &   2.80       & 4           & 12,537        \\ \hline
P19 & 1,086,041& 1,085,897&0.047        & 237.000 & 250.000 &  5.48        & 5           & 15,816        \\ \hline
P20 & 957,372  & 957,319  &0.005        & 228.600 & 248.300 &   8.75       & 56          & 9,035         \\ \hline

\end{tabular}
}      
\vspace{2ex}
\end{table}

\subsection{Comparison to High-Order Masking}

On the surface, HD leaks seem to be a type of second-order leaks,
which involves two values.  For people familiar with high-order
masking~\cite{barthe2015verified}, a natural question is whether the
HD leaks can be mitigated using high-order masking techniques.
To answer the question, we conducted two experiments.  First, we
checked if HD leaks exist in programs equipped with high-order
masking.  Second, we compared the size and execution speed of the code
protected by either high-order masking or our mitigation.

Table~\ref{tbl:comparison} shows the results on benchmarks P4-P9,
which come from Barthe et al.~\cite{barthe2015verified} and thus have
versions protected by $d$-order masking, where $d=2$ to 5.  
While initially we also expected to see no HD leaks in these versions, the
results surprised us.  As shown in the last two columns, HD leaks were
detected in all these high-order masking protected programs.  A closer
look shows that these leaks are all of the $\hdsenSingle{}$ type,
meaning they are due to restriction of the x86 ISA: any binary
operation has to store the result and one of the operands in the same
place, and by default, that place is a general-purpose register.

\begin{table}
\caption{Comparison with order-$d$ masking techniques~\cite{barthe2015verified}.}
\vspace{-2ex}
\label{tbl:comparison}
\centering
\scalebox{0.61}{
\begin{tabular}{|l||c|c|c|c|c|c|}\hline
Name      &Code size (byte) &Run time (us) &HW-leak &HD-leak &\hdsenDouble{} &\hdsenSingle{}       \\\hline
P4 (ours)  &1,171            &0.20         &No       &No       &NONE          &NONE            \\\hline
P4 ($d$=2)     &2,207            &0.75         &No       &Yes      &NONE          &2               \\
P4 ($d$=3)     &4,009            &0.28         &No       &Yes      &NONE          &2               \\
P4 ($d$=4)     &5,578            &0.75         &No       &Yes      &NONE          &2               \\
P4 ($d$=5)     &7,950            &1.00         &No       &Yes      &NONE          &2               \\\hline
P5 (ours)  &1,108            &0.37         &No       &No       &NONE          &NONE            \\\hline
P5 ($d$=2)     &2,074            &0.70         &No       &Yes      &NONE          &2               \\
P5 ($d$=3)     &3,733            &0.60         &No       &Yes      &NONE          &2               \\
P5 ($d$=4)     &5,120            &0.75         &No       &Yes      &NONE          &2               \\
P5 ($d$=5)     &7,197            &0.67         &No       &Yes      &NONE          &2               \\\hline
P6 (ours)  &1,339            &0.27         &No       &No       &NONE          &NONE            \\\hline
P6 ($d$=2)     &3,404            &0.83         &No       &Yes      &NONE          &2               \\
P6 ($d$=3)     &6,089            &0.57         &No       &Yes      &NONE          &2               \\
P6 ($d$=4)     &9,640            &0.80         &No       &Yes      &NONE          &2               \\
P6 ($d$=5)     &14,092           &1.60         &No       &Yes      &NONE          &2               \\\hline
P7 (ours)  &3,223            &0.30         &No       &No       &NONE          &NONE            \\\hline
P7 ($d$=2)     &8,456            &1.41         &No       &Yes      &NONE          &2               \\
P7 ($d$=3)     &15,881           &3.20         &No       &Yes      &NONE          &2               \\
P7 ($d$=4)     &25,521           &4.20         &No       &Yes      &NONE          &2               \\
P7 ($d$=5)     &37,578           &7.80         &No       &Yes      &NONE          &2               \\\hline
P8 (ours)  &3,267            &0.25         &No       &No       &NONE          &NONE            \\\hline
P8 ($d$=2)     &8,782            &1.30         &No       &Yes      &NONE          &2               \\
P8 ($d$=3)     &16,420           &2.00         &No       &Yes      &NONE          &2               \\
P8 ($d$=4)     &26,431           &4.00         &No       &Yes      &NONE          &2               \\
P8 ($d$=5)     &38,996           &8.00         &No       &Yes      &NONE          &2               \\\hline
P9 (ours)      &3,699            &0.45         &No       &No       &NONE      &NONE            \\\hline
P9 ($d$=2) &9,258            &1.15         &No       &Yes      &NONE              &2               \\
P9 ($d$=3)     &17,565           &3.00         &No       &Yes      &NONE          &2               \\
P9 ($d$=4)     &28,189           &5.11         &No       &Yes      &NONE          &2               \\
P9 ($d$=5)     &41,383           &8.40         &No       &Yes      &NONE          &2               \\\hline
\end{tabular}
}
\end{table}

Measured by the compiled code size and speed, our method (which is
already more secure than high-order masking) is more efficient.  In
P9, for example, our mitigated code has 3K bytes in size and runs in
0.45us, whereas the high-order masking protected code has 9K to 41K
bytes (for $d=2$ to 5) and runs in 1.15us to 8.40us.

\subsection{Threat to Validity}

We rely on the HW/HD models~\cite{MangardOP07,mangard2002simple} and
thus our results are valid only when these models are valid.  Although
they have been widely adopted and
validated~\cite{kocher1999differential,clavier2000differential,
brier2004correlation, messerges2000using,moradi2014side}, further
validation is always needed, but it is out of the scope of this work.
We assume the attacker can only measure the power consumption but not
other information such as data-bus or timing.  If such information
becomes available, our mitigation may no longer be secure.

%

Since we focus on cryptographic software, which tends to have simple
program structure and language constructs, there has not been a need
to use more sophisticated points-to analysis than what is provided by
LLVM.  Our analysis is intra-procedural: for cryptographic benchmarks,
we can actually inline all functions to form a monolithic program
before conducting the analysis.
Nevertheless, going forward, these are some of the issues that will be
addressed to broaden the scope of our tool.

\vspace{1mm}
\section{Related Work}
\label{sec:related}

Broadly speaking, existing methods for detecting power side channels
fall into three categories: static analysis, formal verification, and
hybrid approach.  Static analysis relies on compile-time information
to check if masking schemes are implemented
correctly~\cite{barthe2015verified,bayrak2013sleuth,OMHE17,barthe2016strong,BMZ17}.
They are faster than formal verification, which often relies on
SAT/SMT solvers or model counting~\cite{eldib2014formal}.  However,
formal verification is more accurate than static analysis. 
The hybrid
approach~\cite{zhang2018scinfer} aims to combine the two types of
techniques to obtain the best of both worlds.
However, none of these methods focused on the leaks caused by
register reuse inside a compiler, which is our main contribution.

Specifically, although our type based method for detecting 
side-channel leaks is inspired by several prior
works~\cite{OMHE17,barthe2015verified,BMZ17,zhang2018scinfer}, it is
significantly different from theirs.  For example, 
the most recent
method, proposed by Zhang et al.~\cite{zhang2018scinfer}, interleaves
type inference with a  model-counting procedure, with
the goal of detecting HW leaks caused by errors in masking
implementations; however, 
they do not detect HD leaks caused by
register reuse nor remove these leaks.  Their method does not use
Datalog or any of the domain-specific optimizations we have proposed.

Barthe et al.~\cite{barthe2015verified} proposed a relational
technique to check the correctness of masked implementations.
Although the technique covers high-order masking, when applied to a
pair of variables, it has to consider all possible ways in which
second-order leaks may occur, as opposed to the specific type of 
leak involved in register reuse.  Thus, their mitigation would have
been too expensive, in terms of the code size and the execution
speed.  Furthermore, as we have shown in experiments, it does not
seem to be effective in preventing leaks caused by register reuse.



Another difference between our method and existing methods is our
focus on analyzing the word-level representation of a program, as
opposed to a bit-level representation.  While turning a program into a
purely Boolean, circuit-like, representation is
feasible~\cite{zhang2018scinfer, eldib2014formal, almeida2013formal,
bhunia2014hardware}, it does not fit into the standard flow of
compilers.  As such, implementing the approach in compilers is not
straightforward.

The practical security against side-channel leakages via masking can
be evaluated using the ISW model~\cite{IshaiSW03} and subsequent
extensions~\cite{coron2012conversion, balasch2014cost} with
transitions. However, they do not consider leaks that are specific to
register use in modern compilers such as GCC and LLVM.  They do not
consider constraints imposed by the instruction set architecture
either.  Furthermore, they need to double the masking
order~\cite{balasch2014cost} to deal with leaks with transitions, but
still do not eliminate leaks introduced by compilation.

It is known that the security guarantee of software countermeasures
may become invalid after
compilation~\cite{mccann2016elmo,papagiannopoulos2017mind,barthe2018secure,compilerSecure}.
In this context, Barthe et al.~\cite{barthe2018secure} showed that the
compilation process could maintain the constant-time property for
timing side-channel leaks, while our work addresses potential leaks
through power side channels.  Marc~\cite{compilerSecure} also
investigated potential vulnerabilities in power side-channel
countermeasures during compiler optimizations, but did not provide a
systematic method for mitigating them.

Beyond power side channels, there are techniques for analyzing other
types of side channels using logical
reasoning~\cite{ChenFD17,SousaD16,AntonopoulosGHK17}, abstract
interpretation~\cite{DoychevFKMR13,BartheKMO14}, symbolic
execution~\cite{PasareanuPM16,BangAPPB16,PhanBPMB17,BrennanSB18,malacaria2018symbolic} and
dynamic analysis~\cite{WangWLZW17}.
%
%
%
%
As for mitigation, there are techniques that insert masking and other
countermeasures either through
compilers~\cite{BayrakRBSI11,MossOPT12,AgostaBP12,WuGSW18} or through
program synthesis tools~\cite{eldib2014synthesis,BlotYT17}.  However,
these techniques focus exclusively on eliminating the leaks appeared
in the input program.  None of them paid attention to the leaks
introduced by register reuse during the compilation.

\section{Conclusions}
\label{sec:conclusions}

We have presented a method for mitigating a type of side-channel leaks
caused by register reuse in compilers.
The method relies on a type inference system to detect leaks, and
leverages the type information to restrict the compiler's backend to
guarantee that register allocation is secure.
We have implemented the method in LLVM for x86 and evaluated it on
cryptographic software benchmarks. Our experiments demonstrate that the
method is effective in mitigating leaks and the mitigated program has
low performance overhead.
Specifically, it outperforms state-of-the-art high-order masking
techniques in terms of both the code size and the execution speed.

\clearpage
\newpage

\bibliographystyle{ACM-Reference-Format}
\bibliography{hdSCA}

\end{document}